\documentclass[iop]{emulateapj}
\usepackage{graphicx,subfigure,amsmath, amsfonts, amssymb,aas_macros,footnote,color,nnfootnote}
\usepackage[bookmarks=false]{hyperref}
\newcommand{\ha}{H$\alpha$}

\newcommand{\msun}{M$_{\odot}$} 
 
\newcommand{\msunyr}{\msun\ yr$^{-1}$}
\newcommand{\kms}{km~s$^{-1}$}
\newcommand{\ergs}{erg~s$^{-1}$}

\newcommand{\atlas}{A{\small{TLAS}}$^{\rm 3D}$}

\newcommand{\ioutone}{11.93}
\newcommand{\iouttwo}{9.16}
\newcommand{\ioutthree}{7.16}
\newcommand{\fpeaknuc}{24.3}

\newcommand{\fwhmnuc}{0\farcs82}
\newcommand{\fwhmdcnv}{1\farcs05}
\newcommand{\rnuc}{60}

\newcommand{\mnuc}{$4.1\times10^8$}
\newcommand{\signuc}{$2.7\times10^4$}
\newcommand{\nhnuc}{$1.7\times10^{24}$}
\newcommand{\nnuc}{$6.9\times10^3$}
\newcommand{\sfrgas}{3.1}
\newcommand{\mdyn}{$4.9\times10^8$}
\newcommand{\mdmin}{$1.5\times10^8$}
\newcommand{\rdyn}{54}

\newcommand{\inc}{34}
\newcommand{\vrotp}{110}

\newcommand{\mcvc}{$1.1\times10^9$}
\newcommand{\menv}{$7.0\times10^8$}
\newcommand{\rout}{460}
\newcommand{\nout}{$1.0\times10^{16}$}
\newcommand{\mout}{$2.4\times10^7$}
\newcommand{\mhi}{$9.5\times10^6$}
\newcommand{\mhihalf}{$4.8\times10^6$}
\newcommand{\mouttot}{$3.3\times10^7$}
\newcommand{\kinenrg}{$1.0\times10^{55}$}
\newcommand{\kesne}{$10^4$}
\newcommand{\tdyn}{2.6}
\newcommand{\tdynex}{1}
\newcommand{\lumout}{$1.3\times10^{41}$}
\newcommand{\mflux}{13}
\newcommand{\tdepnuc}{32}
\newcommand{\tdep}{85}
\newcommand{\howmany}{3}

\shorttitle{AGN-driven Molecular Outflow from NGC~1266}
\shortauthors{K. Alatalo et al.}
\slugcomment{Accepted by ApJ}

\begin{document}


\title{Discovery of an AGN-driven Molecular Outflow in the Local Early-type Galaxy NGC~1266}

\author{K. Alatalo$^{1}$, 
L. Blitz$^{1}$, 
L.~M. Young$^{2,3}$, 
T.~A. Davis$^{4}$, 
M. Bureau$^{4}$, 
L.~A. Lopez$^{5}$, 
M. Cappellari$^{4}$,
N. Scott$^{4}$,
K.~L. Shapiro$^{6}$,
A.~F. Crocker$^{7}$, 
S. Mart\'{i}n$^{8,9}$,
M. Bois$^{10}$, 
F. Bournaud$^{11}$,  
R.~L. Davies$^{4}$, 
P.~T. de Zeeuw$^{12,13}$, 
P.-A. Duc$^{11}$, 
E. Emsellem$^{10,12}$, 
J. Falc\'{o}n-Barroso$^{14,15}$,
S. Khochfar$^{16}$, 
D. Krajnovi\'{c}$^{12}$, 
H. Kuntschner$^{12}$, 
P.-Y. Lablanche$^{6}$,  
R.~M. McDermid$^{17}$, 
R. Morganti$^{18,19}$, 
T. Naab$^{20}$, 
T. Oosterloo$^{18,19}$, 
M. Sarzi$^{21}$,  
P. Serra$^{18}$, 
A. Weijmans$^{22}$
}


\begin{abstract} We report the discovery of a powerful molecular wind from the nucleus of the non-interacting nearby S0 field galaxy NGC~1266. The single-dish CO profile exhibits emission to $\pm 400$ \kms\ and requires a nested Gaussian fit to be properly described.  Interferometric observations reveal a massive, centrally--concentrated molecular component with a mass of \mcvc\ \msun\ and a molecular outflow with a molecular mass of $ \approx$ \mout\ \msun. The molecular gas close to the systemic velocity consists of a rotating, compact nucleus with a mass of about \mnuc\ \msun\ within a radius of $\approx$ \rnuc\ pc.  This compact molecular nucleus has a surface density of $\approx$ \signuc\ \msun\ pc$^{-2}$, more than two orders of magnitude larger than that of giant molecular clouds in the disk of the Milky Way, and it appears to sit on the Kennicutt-Schmidt relation despite its extreme kinematics and energetic activity.  We interpret this nucleus as a disk that confines the outflowing wind. A mass outflow rate of $\approx$ \mflux\ \msunyr\ leads to a depletion timescale of $\lesssim$ \tdep\ Myr.  The star formation in NGC~1266 is insufficient to drive the outflow, and thus it is likely driven by the active galactic nucleus (AGN).  The concentration of the majority of the molecular gas in the central 100 pc requires an extraordinary loss of angular momentum, but no obvious companion or interacting galaxy is present to enable the transfer.  NGC~1266 is the first known outflowing molecular system that does not show any evidence of a recent interaction.


\end{abstract}


\section{Introduction}
Early-type galaxies (ETGs) are generally thought to be poor in atomic and molecular gas \citep{lees+91}, but their disk galaxy progenitors are generally gas-rich \citep{kauffmann+03}.  How these galaxies transform themselves from gas-rich to gas-poor systems is tied to a host of issues in galaxy evolution, including the migration of galaxies from the blue cloud to the red sequence \citep{Faber+07}, merger-driven starbursts and quasar activity \citep{Hopkins+05}, feedback from star formation and central supermassive black holes on the interstellar medium (ISM) \citep{DiM+05} and radio mode activity \citep{croton+06}.\nnfoottext{\\$^{1}$ Department of Astronomy, Campbell Hall, University of California - Berkeley, California 94720, USA\\
$^{2}$Physics Department, New Mexico Tech, Socorro, New Mexico 87801, USA\\
$^{3}$Adjunct astronomer with the National Radio Astronomy Observatory, NRAO, Socorro, New Mexico 87801, USA\\
$^{4}$Sub-department of Astrophysics, Department of Physics, University of Oxford, Denys Wilkinson Building, Keble Road, Oxford OX1 3RH, UK\\
$^{5}$Department of Astronomy \& Astrophysics, 159 Interdisciplinary Sciences Building, University of California - Santa Cruz, Santa Cruz, California 95064, USA\\
$^{6}$Aerospace Research Laboratories, Northrop Grumman Aerospace Systems, Redondo Beach, CA 90278, USA\\
$^{7}$Department of Astronomy, Lederle Graduate Research Tower B 619E, University of Massachusetts, Amherst, MA 01003, USA\\
$^{8}$European Southern Observatory. Alonso de C\'ordova 3107, Vitacura, Casilla 19001, Santiago 19, Chile\\
$^{9}$Harvard-Smithsonian Center for Astrophysics, Cambridge, MA 02138, USA\\
$^{10}$Universit\'{e} Lyon 1, F-69007; CRAL, Observatoire de Lyon, F-69230 Saint Genis Laval; CNRS, UMR 5574; ENS de Lyon, France\\
$^{11}$Laboratoire AIM, CEA-Saclay/DSM/IRFU/SAp - CNRS - Universit\'{e} Paris Diderot, 91191 Gif-sur-Yvette, France\\
$^{12}$European Southern Observatory, Karl-Schwarzschild-Str 2, 85748 Garching, Germany\\
$^{13}$Sterrewacht Leiden, Leiden University, Postbus 9513, 2300 RA Leiden, the Netherlands\\
$^{14}$Instituto de Astrof\'{i}sica de Canarias, C/ Via Lactea, s/n, E38205 - La Laguna (Tenerife), Spain\\
$^{15}$Depto. Astrof\'{i}sica, Universidad de La Laguna (ULL), E-38206 La Laguna, Tenerife, Spain\\
$^{16}$Max-Planck-Institute for Extraterrestrial Physics, Giessenbachstrae, 85748 Garching, Germany\\
$^{17}$Gemini Observatory, Northern Operations Center, 670 N. A'ohoku Place, Hilo, Hawaii 96720, USA\\
$^{18}$ASTRON, Postbus 2, 7990 AA Dwingeloo, the Netherlands\\
$^{19}$Kapteyn Astronomical Institute, University of Groningen, Postbus 800, 9700 AV Groningen, The Netherlands\\
$^{20}$Max-Planck-Institute for Astrophysics, Karl-Schwarschild-Str.1, 85741 Garching, Germany\\
$^{21}$Center for Astrophysics Research, University of Hertfordshire, Hatfield, Herts AL1 09AB, UK\\
$^{22}$Dunlap Institute for Astronomy and Astrophysics, University of Toronto, 50 St. George St, Toronto, Ontario, M5S 3H4, Canada}

The \atlas\ project aims to address these issues through a complete volume-limited multi-wavelength survey of ETGs within 42 Mpc, spanning a variety of environments and two orders of magnitude in mass \citep{cappellari+11}.  In particular, 259 of the 260 sample galaxies were searched for CO with the Institut de Radioastronomie Millim\'{e}trique (IRAM) 30m telescope \citep{young+11}.  The Combined Array for Research in Millimeter Astronomy (CARMA) interferometer is now mapping the brightest two thirds of the detections in the J=1--0 transition of CO.
Here we report on the S0 field galaxy NGC~1266,  the brightest of the CO detections in the \atlas\ sample.  The distance to NGC~1266 is taken from \atlas, 29.9 Mpc, for which 1\arcsec\ = 145 pc.  The single-dish line profile requires a double Gaussian fit, with broad wings that exceed the escape velocity of NGC~1266.  The interferometric CO observations suggest that NGC~1266 possesses an extraordinarily dense, centrally compact molecular nucleus with a powerful outflow. In \S\ref{obs}, we describe the molecular line observations using the IRAM 30m single-dish, CARMA and Submillimeter Array (SMA) telescopes. We show in \S 3 the two-component structure of the molecular gas -- a massive central compact nucleus and a more extended outflow that appears to be escaping the galaxy, and derive the basic physical properties of the molecular gas.  In \S 4, we compare NGC~1266 to other galaxies and discuss the remarkable properties of its molecular gas. We summarize our main conclusions in \S 5.  

\section{Observations} \label{obs}

NGC~1266 was observed as part of a flux-limited \hbox{CO(1--0)} and \hbox{CO(2--1)} survey of all \atlas\ galaxies carried out at the IRAM 30m telescope \citep{young+11}.  The data consist of a single pointing at the galaxy center, covering a velocity width of 1300 \kms\ with a spectral resolution of 2.6 \kms\ at \hbox{CO(1--0)}, centered on the systemic velocity of 2155 \kms\ as listed in the LEDA catalog (based on absorption lines; \citealt{paturel+03}), with a beamsize of 21\farcs6 in the \hbox{CO(1--0)} line.  The \hbox{CO(2--1)} covered velocity width of 1300 \kms\ with a spectral resolution of 5.2 \kms, with a beamsize of 12\arcsec.  A zeroth order baseline was subtracted, using line-free channels to estimate the continuum contribution to the flux.  NGC~1266 has a peak main beam temperature $T_{\rm mb}$ of 250 mK for \hbox{CO(1--0)} and 630 mK for \hbox{CO(2--1)} \citep{young+11}, or 1.2 Jy for \hbox{CO(1--0)} and 3.5 Jy for \hbox{CO(2--1)} using a conversion factor of 4.73 Jy K$^{-1}$, standard for the IRAM 30m.

NGC~1266 was observed in \hbox{CO(1--0)} with CARMA between 2008 March and 2009 June in three different configurations: D-array (5\arcsec\ resolution at \hbox{CO(1--0)} with robust=0 weighting), B-array (0\farcs7), and A-array (0\farcs3). The primary beam had a diameter of 2\arcmin\ at \hbox{CO(1--0)}, covering the full extent of the optical emission. Due to limited correlator bandwidth, a velocity extent of only 420 \kms\ at 2.5 \kms\ resolution was available for the \hbox{CO(1--0)} line at the time of the observations, enough to adequately measure the properties of the center of the line, but not of the wings.  D-array observations of the bright calibrator \hbox{0423$-$013} were taken in order to correct the slope across the bandpass, followed by alternating integrations on the source and a gain calibrator (0339-107), used to correct atmospheric phase fluctuations.  The CARMA A and B arrays utilize the Paired Antenna Calibration System (PACS; see \citealt{perez+10}), in which the 3.5 m dishes of the Sunyaev-Zel'dovich Array (SZA) are paired with CARMA antennas forming the longest baselines to correct for atmospheric phase fluctuations.  The SZA antennas continually observed a bright quasar (in our case \hbox{0339$-$017}) located 5\fdg9 from the source, which was also used to calibrate the gains.  The flux calibrator was \hbox{0423$-$013}.  At 29.9 Mpc, the highest angular resolution (0\farcs3) corresponds to 44 pc.  The data were reduced in the standard manner, using the Multichannel Image Reconstruction, Image Analysis, and Display ({\tt MIRIAD}) software package \citep{sault+95}.

SMA observations were also obtained between 2008 September and 2009 March in two different configurations: compact (4\arcsec\ resolution at \hbox{CO(2--1)} with robust=0 weighting) and extended (2\arcsec).  The molecular species observed were \hbox{CO(2--1)} and $^{13}$\hbox{CO(2--1)} in the 230 GHz band, and \hbox{CO(3--2)} and HCO$^+$(4--3) in the 345 GHz band.  The primary beam had a diameter of 45\arcsec\ at \hbox{CO(2--1)} and 30\arcsec\ at \hbox{CO(3--2)}, which includes all emission captured by the single dish.  NGC~1266 was imaged with a bandwidth of 2600 \kms\ and a spectral resolution of 2.5 \kms\ at \hbox{CO(2--1)}, centered on the systemic velocity, and a bandwidth of 1700 \kms\ and a spectral resolution of 0.7 \kms\ at \hbox{CO(3--2)}, also centered on the systemic velocity.  We used \hbox{J0423$-$013} and \hbox{J0238+166} as gain calibrators.  The bandpass calibrators were 3C454.3 and 3C273, and the flux was calibrated with Titan and Ganymede.  The zeroth order continuum emission of 40 mJy and 13 mJy at \hbox{CO(3--2)} and \hbox{CO(2--1)}, respectively, was estimated using line-free channels and subtracted from the datacubes.  These continuum levels are consistent with being due to the Rayleigh-Jeans tail of dust continuum.  The data were reduced using the {\tt MIR} package, and analyzed using {\tt MIRIAD}.

\begin{figure*}[t!]\centering
\includegraphics[width=7.2in]{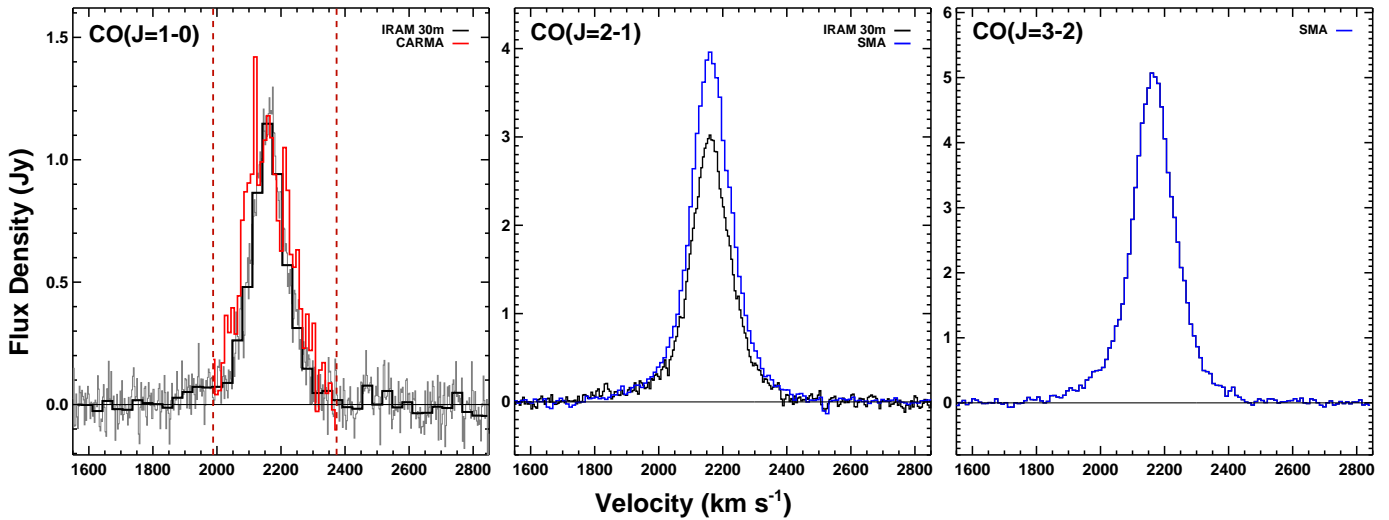}
\caption{{\bf (Left:)} \hbox{CO(1--0)} integrated fluxes from the IRAM 30m (black and gray) and CARMA A, B \& D-arrays (red).  The total CARMA A, B \& D velocity coverage at \hbox{CO(1--0)} is 390 \kms, compared to 1300 \kms\ for the IRAM 30m.  The total velocity widths of both observations are present.  Dashed red lines indicate the CARMA bandwidth.  {\bf (Middle:)} \hbox{CO(2--1)} fluxes from the IRAM 30m (black) and the combined SMA compact (C) and extended (EX) configurations (blue).  {\bf (Right:)}  \hbox{CO(3--2)} flux from the combined SMA C and EX configurations.  CARMA recovers 20\% more flux than the single-dish, and the SMA 30\%, within single-dish baseline and flux calibration uncertainties.}
\label{fig:compflux}
\end{figure*}

NGC~1266 was observed with the Expanded Very Large Array (EVLA) in the D configuration on 2010 March 20 and 28, giving a total of 264 minutes on source with a correlator setup with 256 channels, each of 31.25 kHz width, centered on \ion{H}{1} at the systemic velocity of the galaxy.  The gain calibrator was J0323+0534 and the flux and bandpass calibrator was J0137+3309.  Standard observing and data reduction techniques were used, performed in the Astronomical Image Processing System ({\tt AIPS}) package.  Observations were made at a fixed sky frequency, and the changing topocentric velocity correction between the two observing dates amounts to 0.4 channels; the {\tt AIPS} task CVEL was used to shift one of the datasets before combining them.  The data were Hanning smoothed in velocity, which helps to mitigate the effects of interference at 1408 MHz.   The final, continuum-subtracted image cube was made with robust weighting (robust parameter = 0), has a beam size 66\farcs9$\times$43\farcs0, and covers the velocity range 1321 to 3014 \kms\ at a velocity resolution of 13.4 \kms.  The rms noise level is 0.8 mJy beam$^{-1}$ per channel.   There is no \ion{H}{1} emission evident anywhere within the primary beam (31$'$ FWHM), but prominent \ion{H}{1} absorption against the central continuum source in NGC~1266 is observed.

Archival 1.4 and 5 GHz continuum data obtained at the Very Large Array (VLA) are available for NGC~1266 at resolutions of 1\arcsec\ and 0\farcs4, respectively.  These data for project AB660 were originally taken in December 1992 in the VLA A configuration under the standard $2\times50$ MHz bands with left and right circular polarizations.  Time on source in each frequency was 380 seconds. 3C84 was used as an absolute flux calibrator and B0336--019 was the gain calibrator \citep{bk06}.  The data were reduced in the standard manner using the {\tt AIPS} package.

\section{Results}
\subsection{Molecular Line Profile}
\label{lineprof}

Figure \ref{fig:compflux} presents the integrated spectra of the $^{12}$CO data used in this paper, showing good agreement between the \hbox{CO(1--0)} and \hbox{CO(2--1)} fluxes recovered from the single-dish and the interferometers.  Also of note is that the \hbox{CO(3--2)} line profile from the SMA has the same shape as the \hbox{CO(2--1)}, including wing emission.  The CARMA observations recover $\approx$ 120\% of the single-dish flux, and the SMA observations $\approx 130$\%.  Errors in the single-dish baselines and the intrinsic flux variations of the mm-calibrators can account for the discrepancy.  Therefore, it appears that both CARMA and the SMA recover all of the single-dish flux.

With its narrow single-peaked central velocity component (hereafter CVC) and high-velocity wings, the integrated profile shape is more akin to those observed in protostellar outflows (e.g. \citealt{bl83}) than those of typical external galaxies, where a double-horned profile, is usually seen.


\begin{table}[ht]
\label{tab:gausses}
\begin{center}
\caption{Fitted Gaussian parameters for the \hbox{CO(1--0)}, \hbox{CO(2--1)} and \hbox{CO(3--2)} lines}
\begin{tabular*}{8.5cm}{r l l}
\hline \hline
 & {\bf CVC} & {\bf broad wing}\\
 \hline
{\bf \hbox{CO(1--0)}} & & \\
$v_{\rm sys} = $& $2162\pm1.4$ \kms & $2142\pm20$ \kms \\
${\rm FWHM}  = $& $114\pm3.9$ \kms & $353\pm17$ \kms \\
$T_{\rm mb}{\rm (peak)} = $& $0.21\pm0.006$ K & $0.03\pm0.005$ K \\
I$_{\rm CO} = $& $25.5\pm1.1$ K \kms & $9.6\pm1.8$ K \kms \\
\hline
{\bf \hbox{CO(2--1)}} & & \\
$v_{\rm sys} = $& $2161\pm0.5$ \kms & $2151\pm4.4$ \kms \\
${\rm FWHM}  = $& $128\pm1.9$ \kms & $353\pm17$ \kms \\
$T_{\rm mb}{\rm (peak)} = $& $0.53\pm0.01$ K & $0.10\pm0.01$ K \\
I$_{\rm CO} = $& $71.1\pm1.6$ K \kms & $36.7\pm3.8$ K \kms \\
\hline
{\bf \hbox{CO(3--2)}} & & \\
$v_{\rm sys} = $& $2166\pm0.4$ \kms & $2162\pm3.7$ \kms \\
${\rm FWHM}  = $& $134\pm1.3$ \kms & $353\pm17$ \kms \\
$F_{\rm peak} = $& $4.15\pm0.03$ Jy & $0.82\pm0.03$ Jy \\
S$_{\rm CO} = $& $589\pm7.5$ Jy \kms & $306\pm18$ Jy \kms \\
\hline \hline
\end{tabular*} \\
\vskip 1mm \end{center}
{\bf Note:} I$_{\rm CO}$ is the integrated CO line intensity from the IRAM 30m, similarly for S$_{\rm CO}$ and the SMA.  The broad wing component accounts for 27\% and 34\% of the total flux for \hbox{CO(1--0)} and \hbox{CO(2--1)}, respectively.  The broad wing component of \hbox{CO(3--2)} also contributes 34\% of the total flux.   The 7\% difference in the broad wing contribution between the transitions is attributed to the uncertainties in the fits to the line profiles.  We will adopt 34\% as the more likely value, due to the better signal-to-noise ratios at \hbox{CO(2--1)} and \hbox{CO(3--2)}.  \hbox{CO(3--2)} data were only obtained with an interferometer, thus the natural unit of this line Jy.  To convert the \hbox{CO(3--2)} fluxes to Kelvin, we first divide by 9 (the ratio of the \hbox{CO(3--2)} and \hbox{CO(1--0)} frequencies squared), then multiply by the K per Jy conversion factor of the adopted beam of 4.73 Jy K$^{-1}$.  The errors listed are the formal uncertainties of the fits.  The value and uncertainty of the FWHM of the broad wing component for all transitions are adopted from the fit to the \hbox{CO(2--1)} line.
\end{table}

\begin{figure}[h] \centering
\includegraphics[width=2.6in,clip,trim=0cm 0cm 0cm 0cm]{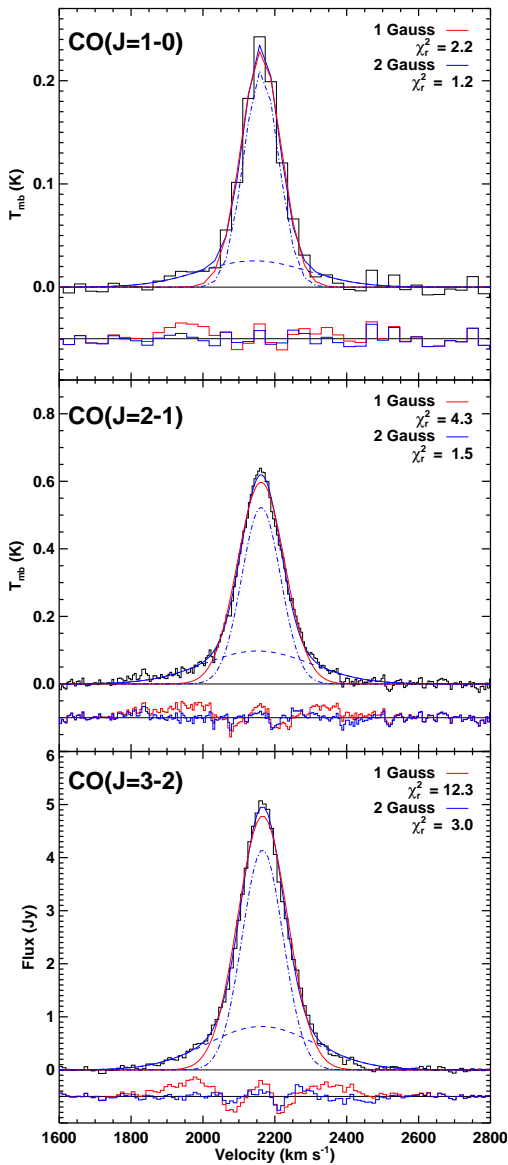}
\caption{\hbox{CO(1--0)} (top), \hbox{CO(2--1)} (middle) and \hbox{CO(3--2)} (bottom) integrated profiles of NGC~1266 from the IRAM 30m and SMA (\hbox{CO(3--2)} only). We also show a comparison between a one Gaussian fit the the data (red) and a two-Gaussian fit (blue) consisting of a narrow (blue dot-dashed line) and a broad (blue dashed line) component.  Residuals are plotted below the fits and show that a two-Gaussian fit is a much better match to the data.  The width of the broad component is fixed to 353 \kms, initially constrained using the fit the to the \hbox{CO(2--1)} line.  The goodness of fit improves significantly with the addition of the broad component, illustrated both in the measured $\chi^2$ as well as in the residuals.} 
\label{fig:2gauss} 
\end{figure}

\begin{figure}[ht] \centering
\includegraphics[width=3.5in,clip,trim=1cm 0cm .6cm 0cm]{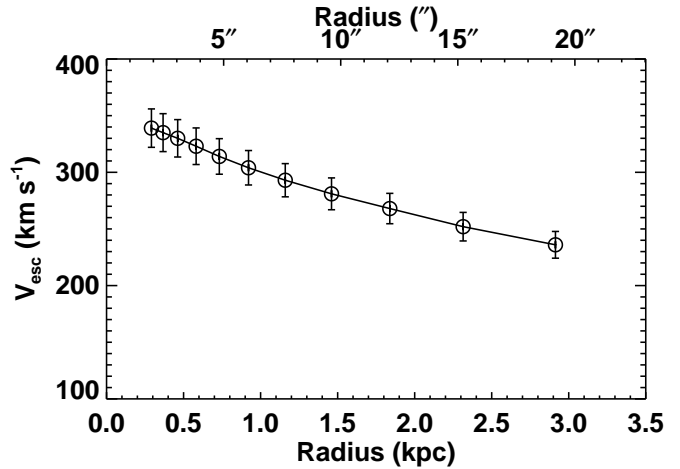}
\caption{The radial escape velocity profile of NGC~1266.  The escape velocity was derived using Jeans Anisotropic Multi-Gaussian Expansion modeling (JAM)  of the stellar kinematics (\citealt{scott+09}, 2010, in prep).}
\label{fig:vesc}
\end{figure}

Fig. \ref{fig:2gauss} shows the fits and residuals of the single and nested Gaussians.  Table 1 gives a summary of the fitted Gaussian parameters.  We find that the broad wing and CVC, which we identify respectively with the broad and narrow Gaussian of the nested Gaussian fits, encompass respectively 34\% and 66\% of the \hbox{CO(2--1)} single-dish integrated flux, which has the best signal-to-noise ratio.  When comparing the line to the escape velocity radial profile measured in the plane of the galaxy (see Figure \ref{fig:vesc}), we find that at least 2.5\% of the wing emission is from velocities exceeding the innermost local escape velocity (calculated at 2\arcsec) of $v_{\rm esc} \approx 340$ \kms.  The true percentage may be higher, depending on the geometry of the high velocity gas.  If the molecular gas is located either out of the plane of the galaxy or at larger radii, the local potential (and thus $v_{\rm esc}$) would be much reduced.  The $v_{\rm esc}$ profile was calculated using Jeans Anisotropic Multi-Gaussian Expansion (MGE) modeling (JAM; \S\ref{core}), to extract the true potential of the galaxy by modeling the observed stellar kinematics (\citealt{scott+09}, in prep).

\subsection{Molecular Gas Mass}
\label{molextent}

Using the CO (1--0) single-dish flux quoted in Young et al. (2011), a column density of molecular hydrogen N(H$_2$) = $X_{\rm CO}\int T_{\rm mb}~\mathrm{d}v$, where the conversion factor $X_{\rm CO} = 2\times10^{20}~{\rm cm}^{-2}$ (K km s$^{-1})^{-1}$ \citep{dame+01}, a beamsize of 21\farcs6, and a beam area of 529 sq.arcsec (which includes a beamshape correction of 1/ln(2); \citealt{baars07}), the total mass of the molecular gas would be $1.7\times10^9$ \msun, including a correction factor of 1.36 for He.  However, the wing emission is optically thin (see \S\ref{wings}) and contributes little to the total mass.  Considering only the CVC, the molecular gas mass is \mcvc\ \msun, including He. 

\begin{figure*}[t]
\centering
\includegraphics[width=7in,clip,trim=0cm 0cm 1cm 0.8cm]{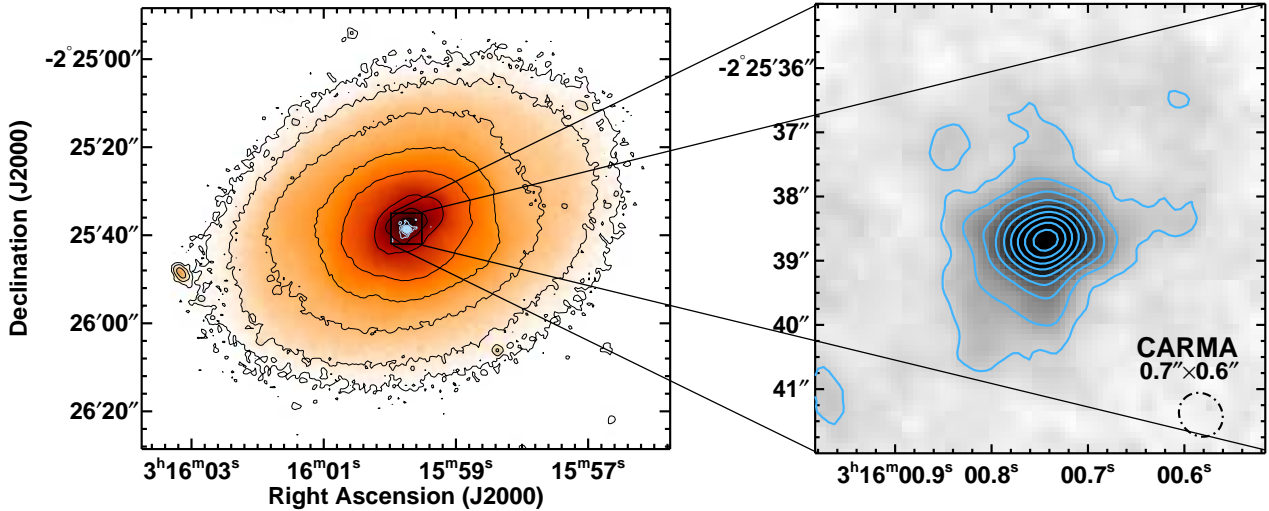}
\caption{{\bf(Left)} Contours of the integrated \hbox{CO(1--0)} molecular gas detected by CARMA (blue), overlaid with the isophotes (black) and grayscale of an R-band image from the Cerro Tololo Inter-american Observatory (CTIO) 1.4m telescope (SINGS). {\bf (Right)} A zoomed-in version of the \hbox{CO(1--0)} integrated intensity map.  The synthesized beam (0\farcs7$\times$0\farcs6) is plotted in the lower right corner.  The CARMA \hbox{CO(1--0)} integrated intensity map was created by summing the full velocity width of the CARMA cube, including the A, B and D arrays.  Robust = 0 weighting was used in order to reveal low surface brightness, more diffuse structures at the expense of resolving the most compact components.  Contour levels are 3, 6, 9, 12, 15, 18, 21, 24 and 27 Jy beam$^{-1}$ \kms\ (rms = 1.0 Jy beam$^{-1}$ \kms).}
\label{fig:co_extent}
\end{figure*}

\citet{draine+07} derive a total dust mass of 1.02 $\times\ 10^7$ \msun\ in NGC~1266 (rescaled to our distance of 29.9 Mpc), using {\it Spitzer} Infrared Nearby Galaxy Survey (SINGS) data.  Since a significant \ion{H}{1} reservoir is not observed in NGC~1266 (see below), we may assume that all of the measured dust is associated with the molecular gas.  Adopting a gas-to-dust mass ratio of 100 \citep{sav+math79}, the dust mass implies a molecular gas mass of $1.0 \times\ 10^9$ \msun.  The derived dust mass is thus in good agreement with the mass derived from the CO measurements, and a standard $X_{\rm CO}$ conversion factor.

\subsection{Core of the Emission}
\label{core}

Figure \ref{fig:co_extent} shows the full spatial extent of the CARMA \hbox{CO(1--0)} integrated emission ($\Delta v = 390$ \kms), overlaid on an R-band image of the galaxy from SINGS, illustrating that the vast majority of the molecular gas is restricted to the innermost region of NGC~1266.  The righthand panel of the figure shows that the molecular gas is not well described by a single spatial component, but instead as both a diffuse, relatively extended component, which we will call the ``envelope,'' and a highly concentrated nuclear structure, which we will call the ``nucleus''.

\begin{figure}[hb]
\centering
\includegraphics[height=3.2in,clip,trim=0.5cm 1.1cm 1.3cm 1.3cm,angle=90]{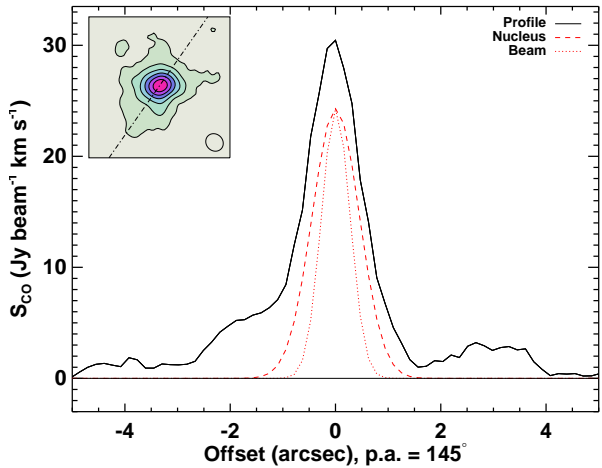}
\caption{Slice of the \hbox{CO(1--0)} integrated intensity map from CARMA at a 145$^\circ$ position angle (see inset).  The radial profile requires two Gaussians for a good fit, an ``envelope'' and a ``nuclear'' component.  The nuclear component (red dashed line) has a peak intensity of \fpeaknuc\ Jy beam$^{-1}$ km s$^{-1}$ and a FWHM of 1.05\arcsec, (\fwhmnuc\ after correcting for convolution effects from the beam). The radial profile of the beam is also shown (red dotted line), to illustrate that the nucleus is spatially resolved.}
\label{fig:coreslice}
\end{figure}

In order to separate the nucleus from the envelope in the CARMA map, we take a slice across the integrated intensity map at a position angle of 145\degr, to trace the elongation of the narrow component, and fit two Gaussians, to the nucleus and the envelope, respectively.  Figure \ref{fig:coreslice} illustrates the radial cut as well as the the nuclear Gaussian component, which has a characteristic FWHM of \fwhmdcnv.  We use FWHM/2 to define a characteristic deconvolved radius \hbox{$R_{\rm nuc}$ of $\approx $ \rnuc\ pc} (after accounting for the CARMA beam).  Using this, the peak intensity of the profile in Fig. \ref{fig:coreslice} (\fpeaknuc\ Jy beam$^{-1}$ km s$^{-1}$), the pixel size of 0\farcs08 pix$^{-1}$, and $X_{\rm CO} = 2\times10^{20}$ cm$^{-2}$ (K km s$^{-1}$)$^{-1}$, we derive a molecular gas mass for the nucleus of $M_{\rm nuc, H_2+He} =$ \mnuc\ \msun, including He.  As the spatial distribution of the broad wing emission in the central region of the CARMA \hbox{CO(1--0)} integrated image is unknown, we do not subtract a potential wing contribution from this nucleus molecular gas mass.  If we assume that the rest of the mass in the CVC comes from the envelope, we deduce that $M_{\rm env} \approx$ \menv\ \msun, including He.

\begin{figure}[hb]
\centering
\includegraphics[width=3.4in,clip,trim=0cm 2cm 0cm 3.5cm]{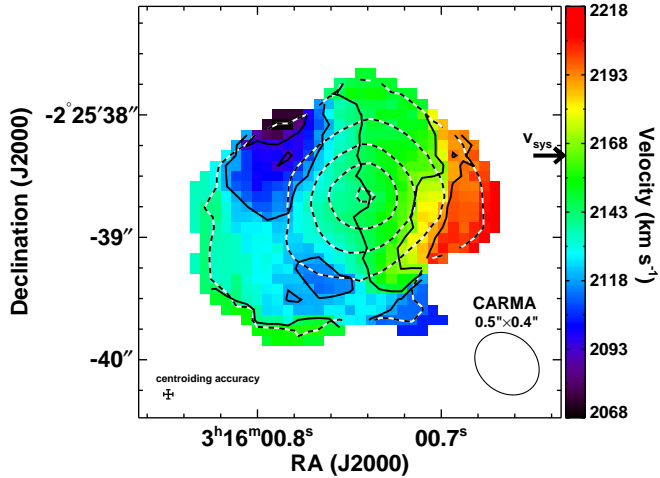}
\caption{\hbox{CO(1--0)} mean velocity map from CARMA overlaid with contours from the corresponding integrated intensity map (black and white).  A velocity gradient is clearly seen across the nucleus, interpreted as a thick rotating disk (see text).  These velocity moments were constructed from the CARMA A and B array data only (synthesized beam 0\farcs5$\times$0\farcs4), to extract structures at the smallest spatial scales at the expense of the extended, diffuse emission.  The velocity contours are spaced at 25\kms, or 2069, 2094, 2119, 2144, 2169, and 2194 \kms.  The systemic velocity of 2160 \kms\ is denoted with an arrow.  The centroiding accuracy is derived using the FWHM$_{\rm beam}$/2 SNR \citep{vlassII}, where SNR is the signal-to-noise ratio and is at least 3 per channel, yielding an accuracy of at least 0\farcs08.}
\label{fig:mom1}
\end{figure}

\begin{figure}[ht!]
\centering
\includegraphics[width=2.2in,clip,trim=1.2cm 0.7cm 1.5cm 1.6cm]{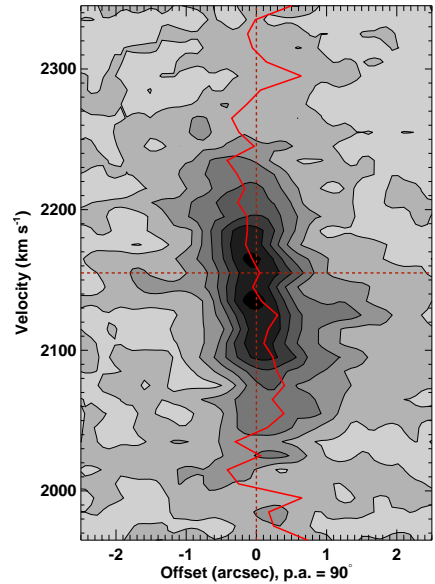}
\caption{Position-velocity diagram of NGC~1266 for our \hbox{CO(1--0)} A and B array CARMA data, taken at position angle of 90\degr.  The observed lack of a turnover is indicative of the gas not reaching the flat part of the rotation curve.  The data are consistent with rotation with a gradient of 1.8 \kms\ pc$^{-1}$ in the inner parts.  The red line is a trace of the peak at each velocity.  The dark red dashed lines indicate $v_{\rm sys}$ and the center.}
\label{fig:co_pv}
\end{figure}

Figure \ref{fig:mom1} is a map of the first moment of the \hbox{CO(1--0)} emission, and shows a clear velocity gradient across the nucleus.  Figure \ref{fig:co_pv} is the position-velocity (PV) diagram at a position angle of 90\degr\ (i.e., roughly along the kinematic major axis), suggestive of a rotating disk with a velocity gradient of 1.8 \kms\ pc$^{-1}$.  Rotation can be traced on the PV diagram to a radius of \rdyn\ pc, consistent with the radius of the nucleus derived above.  At $R_{\rm nuc} \approx$ \rnuc\ pc, we derive a maximum projected rotation velocity of the nucleus of $v_{\rm rot} \approx$\vrotp\ \kms.  The corresponding enclosed dynamical mass, assuming circular rotation and spherical symmetry ($M_{\rm dyn,nuc} = R_{\rm nuc} v_{\rm rot}^2/\sin^2i_{\rm nuc} G$, where $i_{\rm nuc}$ is the inclination of the nuclear disk and $G$ is the gravitational constant), is uncertain because of the unknown inclination, but the formal lower limit ($i_{\rm nuc} = 90^\circ$) is \mdmin\ \msun.  A rough inclination estimate based on the axial ratio of the \hbox{CO(1--0)} integrated intensity map shown in Figure \ref{fig:co_extent} and the assumption of a thin disk yields $i_{\rm nuc} \approx$ \inc\degr, and thus an enclosed dynamical mass \hbox{$M_{\rm dyn,nuc} \approx $ \mdyn\ \msun}.  The dynamical mass calculated using the rotation of the nucleus is therefore consistent with the $^{12}$CO-derived molecular gas mass of the nucleus, when taking into account a correction associated with the inclination.  The stellar contribution to the nucleus mass is discussed later.  

An independent estimate of the mass of molecular gas in the core was obtained via dynamical modeling of the stellar kinematics. For this we constructed a MGE model \citet{emsellem94} of the stellar mass distribution of NGC~1266 from our own R-band Isaac Newton Telescope (INT) photometry (Scott et al. 2011, in prep) and the software of \citet{capp02}. We used this MGE model to build an axisymmetric dynamical model of the stellar kinematics based on the JAM formalism \citep{capp08}. The global anisotropy, inclination and mass-to-light ratio (M/L) of the galaxy were fitted to the large-scale \atlas\ {\tt SAURON} stellar kinematics (Cappellari et al. 2011b, in prep) and will be discussed elsewhere. In this particular case the JAM model also explicitly includes as an extra free parameter: the mass of the molecular gas in the core, modeled as a dark Gaussian with a width of 1\farcs2. The mass of molecular gas that best fits the {\tt SAURON} stellar kinematics in the centre is $M_{\rm gas} \approx 8 \times 10^8$ \msun, with a factor $\approx 3$ uncertainty dominated by systematic effects. This value is an order of magnitude larger than the mass in stars within 100 pc of the center, $M_{\rm \star,100pc} \approx 5\times10^7$ $M_\odot$, as determined from an analogous MGE mass model without the molecular core. This result is thus consistent with the mass in the center of NGC~1266 being dominated by molecular gas.

\begin{figure}[h!]
   \centering
\includegraphics[width=3.4in,clip,trim=1.3cm 0.5cm 0cm 1cm]{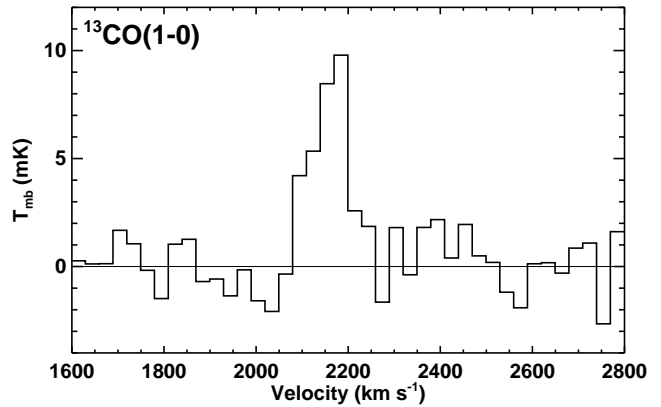}
\caption{$^{13}$CO(1-0) spectrum taken with the IRAM 30m telescope, as part of a larger census of dense gas in \atlas\ ETGs (Crocker et al. 2011, in prep). The beam at this frequency has a 22\arcsec\ half-power beam width (HPBW) and the channels shown are 30 km s$^{-1}$. } 
   \label{fig:13co}
\end{figure}

One final consistency check on the molecular gas mass can be performed using the \hbox{$^{13}$CO(1--0)} single-dish profile obtained as part of the dense gas census of \atlas\ galaxies (Crocker et al. 2011, in prep), shown in Fig. \ref{fig:13co}.  The total \hbox{$^{13}$CO(1--0)} line intensity is 0.98 K \kms\ within the 22\farcs4 beam of the IRAM 30m telescope.  We must first correct $I(^{13}{\rm CO})$ to reflect a more realistic size for the emitting region, and adopt a radius of $\approx$ 1\arcsec.  The correction corresponds to the ratio of the two solid angles subtended: $\Omega_{\rm beam} / \pi R^2$, in this case 567 arcsec$^2$/3.14 arcsec$^2$.  The resulting corrected $I(^{13}{\rm CO})$ is 177 K \kms.  Using the \hbox{$^{13}$CO} to $A_V$ conversion from \citet{lada2}, \hbox{$I(^{13}{\rm CO})/({\rm K~km~s^{-1}) = 1.88 + 0.72~mag^{-1}}~A_V $} results in an extinction $A_V \approx 243$ mag.  Converting $A_{\rm V}$ to N(H$_2$) using $N({\rm H_2})/A_V = 9\times10^{20}$ cm$^{-2}$ mag$^{-1}$ from \citet{schultz+75} and \citet{bohlin+78}, and then converting N(H$_2$) to a molecular gas mass, yields $M_{\rm nuc} \approx$ \hbox{$3.2 \times10^8$ \msun}, corrected for He.  Therefore the $^{13}$CO--derived mass is in reasonable agreement with the \hbox{$^{12}$CO}-derived molecular gas mass.
 
With a molecular gas mass of \mnuc\ \msun\ (including He), assuming a circular disk, the mean molecular hydrogen surface density in the nucleus is $\Sigma_{\rm H_2} \approx$ \signuc\ \msun\ pc$^{-2}$ (corresponding to a mean column density $N({\rm H_2}) \approx$ \nhnuc\ cm$^{-2}$) and the mean volume density of H$_2$ is at least \nnuc\ cm$^{-3}$ (lower limit assuming spherical symmetry).  Given that tracers with higher critical densities, such as $^{13}$CO, CS, HCO$^+$ and HCN, are robustly detected in the nucleus (Alatalo et al. 2011, in prep), the molecular gas is likely to be clumpy.


The dynamical mass, JAM mass and $^{13}$CO-derived mass agree quite well with the \hbox{$^{12}$CO(1--0)}-derived molecular gas mass of the nucleus, indicating that using the standard Milky Way conversion factor $X_{\rm CO} = 2\times10^{20}~{\rm cm^{-2}~(K~km~s^{-1})^{-1}}$ is appropriate to determine the mass of the molecular material in the center of NGC~1266.  The JAM modeling also indicates that the stellar contribution to the mass of the region with molecular gas is negligible. 

\subsection{Wings of the Emission}
\label{wings}

\begin{figure}[ht!]
   \raggedleft
	\subfigure{\includegraphics[height=3.0in,angle=90,clip,trim=0cm 1.3cm 0cm 1.3cm]{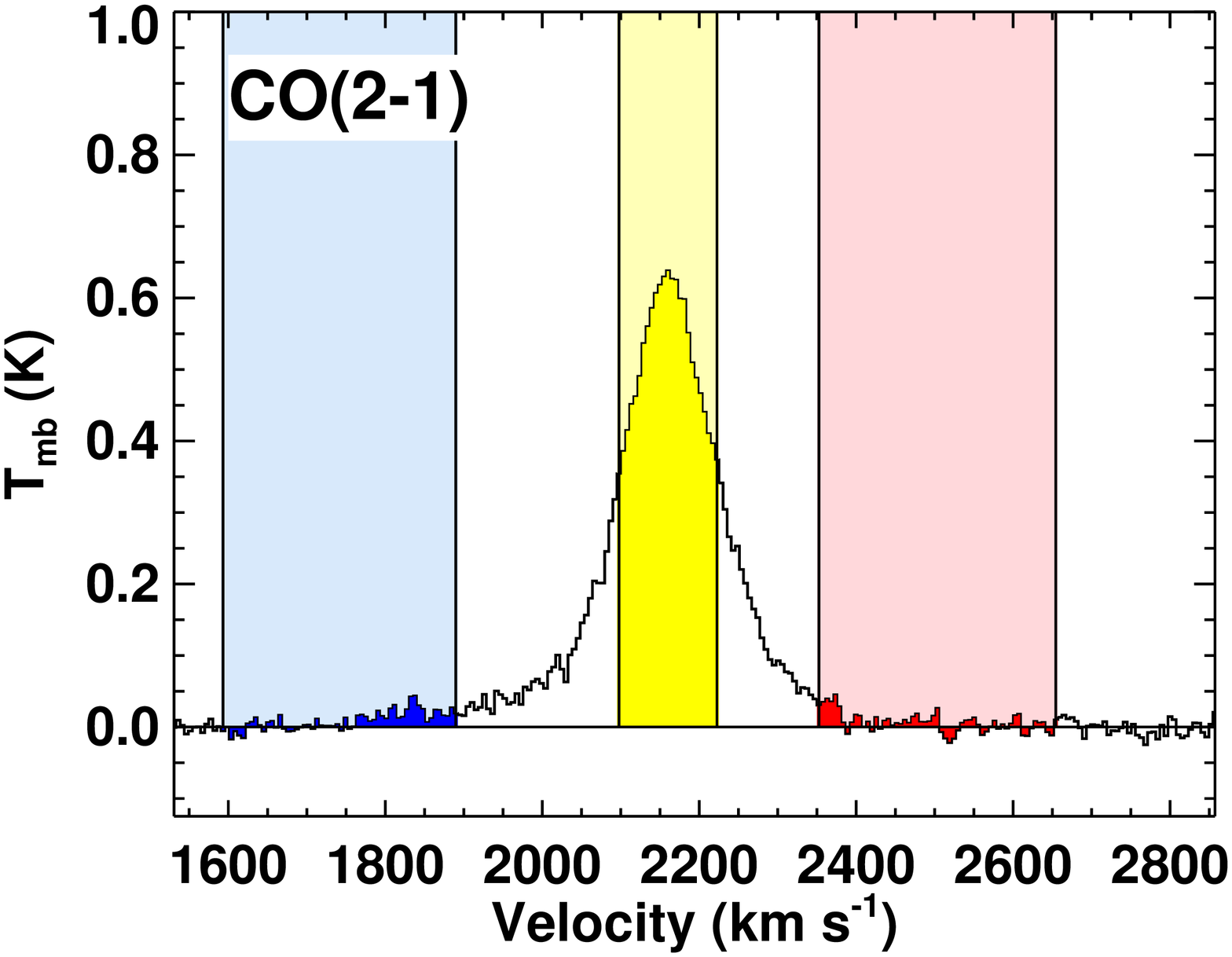} }\vskip -3mm
	\subfigure{\includegraphics[width=3.3in,clip,trim=0.4cm 0.6cm 1.4cm 0.6cm]{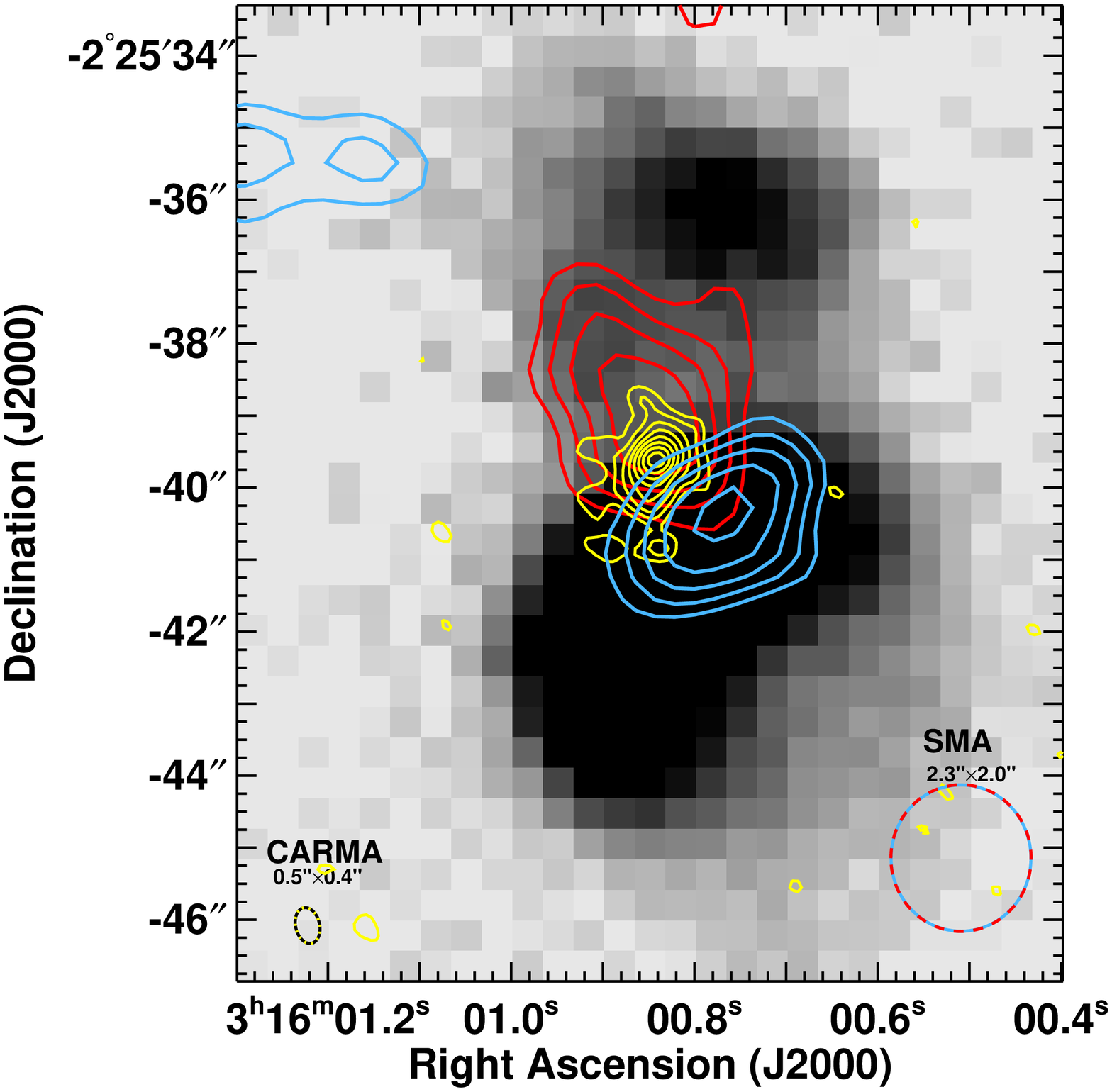} }
	\caption{{\bf (Top:)} \hbox{CO(2--1)} spectrum from the IRAM 30m telescope, indicating the velocity ranges used.  {\bf (Bottom:)} CO core and wings in NGC~1266, overlaid on a greyscale \ha\ narrow-band image from SINGS.  Superimposed contours are from the CARMA \hbox{CO(1--0)} integrated intensity map (yellow) and the SMA \hbox{CO(2--1)} redshifted (red) and blueshifted (blue) wings.  Contours are 3, 4, 5 and 6 Jy beam$^{-1}$ \kms\ for the red and blue components (rms = 1 Jy beam$^{-1}$ \kms) and 6.4, 8.6, 10.8, 13.0, 15.2 and 17.4 Jy beam$^{-1}$ \kms\ for the yellow component (rms = 1.1 Jy beam$^{-1}$ \kms)  }
   \label{fig:co_mom1}
\end{figure}

We use the \hbox{CO(2--1)} SMA data to resolve the spatial extent of the high velocity wings of the molecular gas, because of the limited velocity coverage of the correlator when the CARMA data were taken. In order to minimize contamination from the nucleus, we made a map of the emission over the velocity ranges $v$=1590--1890 \kms\ and $v$=2350--2650 \kms, eliminating the central 460 \kms.   Figure \ref{fig:co_mom1}a shows the velocity ranges of the CVC and the wings of the line used for mapping.  The offsets from $v_{\rm sys}$ were chosen to achieve equal signal-to-noise ratios of 6 for the peaks of the red- and blue-shifted images in Figure \ref{fig:co_mom1}b, which shows that the wings are considerably more extended spatially than the nucleus.  If we model the outflow as a bi-spherical structure, similar to what is observed in M82 \citep{walter+02L}, and represent the red- and blueshifted lobes by two tangent spheres whose point of contact is the nucleus, the distance the molecular gas has traveled from the center is equal to the width of the lobes.  We measure the width of the blueshifted and redshifted lobes to be $\approx$ 3\farcs4 and 4\farcs2, respectively.  Taking the average of the lobe extents, and correcting for convolution with the 2\arcsec\ SMA beam gives us a characteristic extent of $\approx$ 3\farcs2, or \rout\ pc.  The red- and blue-shifted wings are also spatially offset from one another.  The position centroids of the lobes are separated by $\approx$ 2\farcs3, thus each is offset by about 170 pc from the galaxy center.  The large spatial extent of the wing emission rules out an interpretation whereby the high gas velocities are the result of gas having fallen into the central potential well of the galaxy.  We thus conclude that the source of the high velocity wing emission is a kpc--scale molecular outflow, with $R_{\rm outflow} \approx$ \rout\ pc.

Further evidence that the wings represent an outflow is that the axis connecting the lobes is intersected by the CVC and is not aligned with the kinematic major-axis of the nuclear gas disk.  This suggests that the broad wing emission is composed of gas that is being expelled, and is not an extension of the nuclear material. If we estimate the inclination of the outflow (with respect to the plane of the sky) using the average offset of the centroids with respect to the nucleus (2\farcs3) divided by the average extent of the lobes (3\farcs2) calculated above, we obtain an inclination angle of roughly 20\degr.

The mass of outflowing gas is somewhat uncertain because the wings are unambiguously detected only in $^{12}$CO (Fig. \ref{fig:compflux}), requiring that the density, optical depth and temperature of the outflow be determined from the excitation of the three lowest CO rotational transitions.  We use the beam-corrected (to the single-dish \hbox{CO(1--0)} beam of 21\farcs6) intensities of the wing emission from the \hbox{CO(1--0)} and (2--1) IRAM 30m single-dish spectra and the \hbox{CO(3--2)} SMA spectrum (see Table 1).  The beam-corrected values are \ioutone\ K \kms\ for \hbox{CO(1--0)}, \iouttwo\ K \kms\ for \hbox{CO(2--1)} and \ioutthree\ K \kms\ for \hbox{CO(3--2)}.

We use the {\tt RADEX} large velocity gradient (LVG) software \citep{vandertak+07} to determine volume densities and column densities in the wind.  With {\tt RADEX}, assuming a line width $v_{\rm fwhm} = 353$ \kms, we find that the conditions required to reproduce the beam-corrected flux ratios above are an H$_2$ density $n({\rm H_2}) \approx 10^3~{\rm cm}^{-3}$, a kinetic temperature $T_{\rm kin} \approx 100$ K, and a CO column density $N({\rm CO}) \approx$ \nout\ cm$^{-2}$.  The $N$(CO) was also constrained using on the optically thin estimate of \citet{knapp+jura76} assuming subthermally excited CO, and the result is consistent with that from {\tt RADEX}.  Assuming a CO/H$_2$ abundance ratio of 10$^{-4}$, the derived $N$(CO) and our adopted beamsize of 21\farcs6 imply a molecular outflow mass of \mout\ \msun, including He.  It is however possible that the outflowing molecular mass is higher.  The optically thin approximation (i.e. that we are seeing all CO molecules) provides us with a lower limit to the total outflowing molecular mass.  If the state of the molecular gas in the outflow of NGC~1266 is more akin to gas in either ULIRGs or giant molecular clouds (GMCs) in the Milky Way, where a conversion factor ($X_{\rm CO}$) is used, the total outflowing mass could be a factor of \hbox{$\approx$ 10 -- 20} higher.

\subsection{Additional Data}
\label{ancillary}

\begin{figure}[h!]
\centering
\subfigure{\includegraphics[height=3.0in,clip,trim=0.4cm 1.1cm 1.1cm 0.5cm,angle=90]{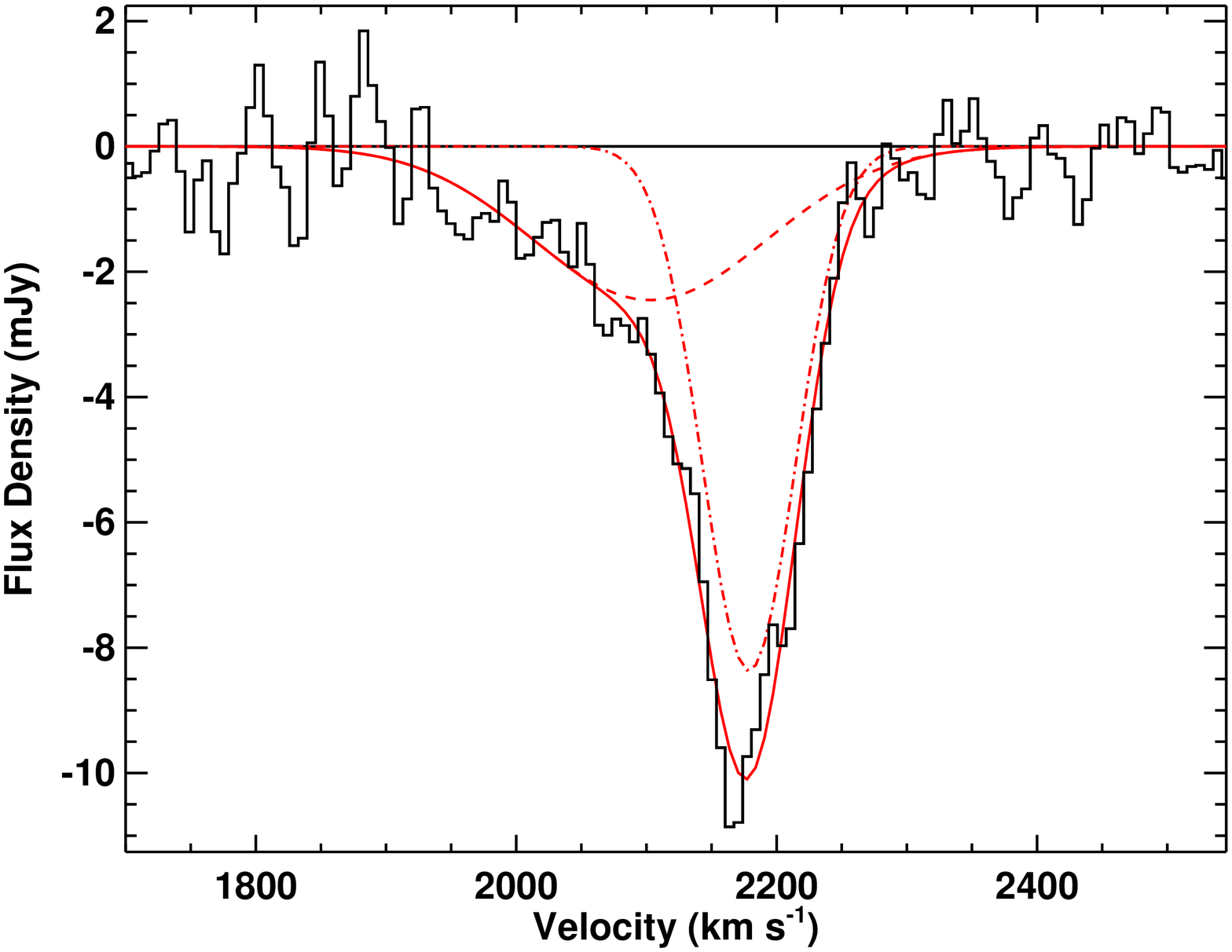}}
\subfigure{\includegraphics[height=3.0in,clip,trim=0.7cm 1.2cm 1.2cm 0.5cm,angle=90]{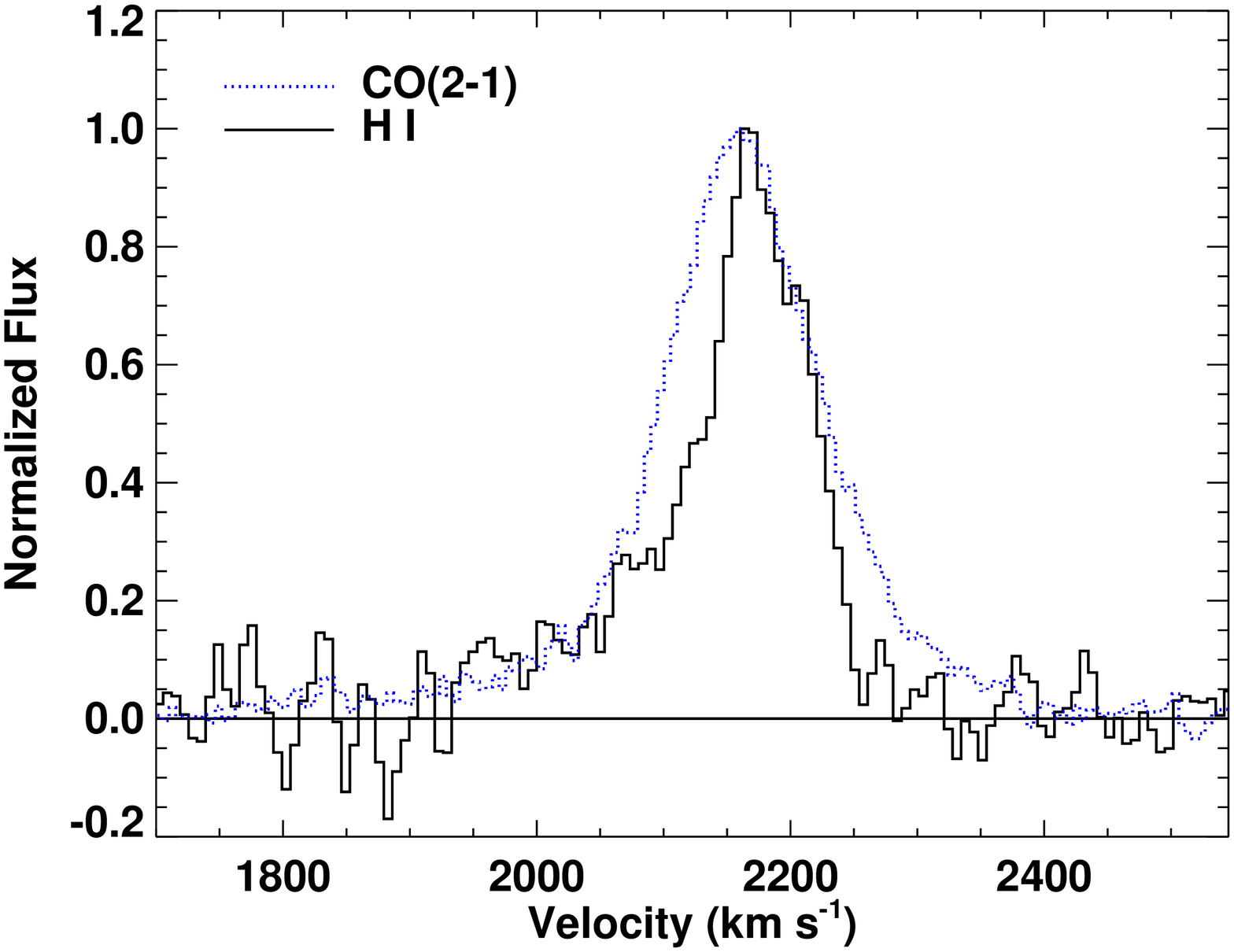} }
\caption{{\bf (Top:)} Continuum-subtracted \ion{H}{1} profile in the central pixel of the unresolved map observed with the EVLA D-array (black).  The continuum level for the source is 113 mJy.  The blueshifted wing is evident in the absorption profile, reaching $\gtrsim 200$ \kms\ before dropping into the noise.  Overlaid (red) is a two-Gaussian fit to the absorption profile, including a blueshifted high-velocity contribution (dashed) and a deeper absorption contribution near $v_{\rm sys}$ (dot-dashed).  {\bf (Bottom:)}  A direct comparison between the \ion{H}{1} absorption profile (black) and the \hbox{CO(2--1)} emission spectrum from the IRAM 30m telescope (blue dotted).  The CO emission and \ion{H}{1} absorption profiles trace each other well at high blueshifted velocities, and the marked absence of redshifted \ion{H}{1} absorption is confirmation that NGC~1266 harbors an \ion{H}{1} outflow.}
\label{fig:co_hi}
\end{figure}

While no \ion{H}{1} emission is detected in NGC~1266, \ion{H}{1} appears in absorption against the 1.4 GHz continuum source.  The \ion{H}{1} observations have thus far only been taken at low resolution, so are presently unresolved.  The top panel of Figure \ref{fig:co_hi} shows the \ion{H}{1} absorption profile, which exhibits a broad blueshifted velocity component, and the fitted Gaussians describing the absorption profile.  The bottom panel of Figure \ref{fig:co_hi} shows a good correspondence between the blueshifted absorption feature in the \ion{H}{1} and the high-velocity wing of the \hbox{CO(2--1)} emission line profile.  This identifies the blueshifted molecular gas as moving out (as opposed to infalling from behind), and implies that the outflow in NGC~1266 has multiple phases.

Calculating the column density from the absorption feature, assuming $T_{\rm spin} = 100$ K, reveals that the \ion{H}{1} column in front of the continuum source is $N_{\rm H} = 2.1\times10^{21}~{\rm cm}^{-2}$, with the outflowing component contributing $8.9\times10^{20}$ cm$^{-2}$ and the systemic component contributing $1.2\times10^{21}~{\rm cm}^{-2}$.  If we assume that the \ion{H}{1} and CO are co-spatial ($R_{\rm H I} = R_{\rm outflow} =$ \rout\ pc), and that the total mass of outflowing \ion{H}{1} is twice what we calculate (since we are only able to detect the blueshifted lobe), we derive the total \ion{H}{1} mass in the outflow to be \mhihalf\ \msun\ per lobe, totaling \mhi\ \msun.  If we include the \ion{H}{1} contribution, the total neutral gas mass (\ion{H}{1} + H$_2$ + He) of the outflow is thus \mouttot\ \msun.

The molecular gas (as traced by CO) and atomic gas are likely not the only constituents of the mass in the outflow.  By not accounting for other states of the gas, we are underestimating the true outflow mass.  For example, \citet{roussel+07} detect a large reservoir of warm H$_2$ in NGC~1266 using {\it Spitzer}, with a total mass of $M_{\rm H_2,warm} \approx 1.3\times10^7$ \msun.  It is very likely that this mass of warm H$_2$ belongs to the outflow rather than the nuclear disk.  The total warm H$_2$ luminosity is $L_{\rm H_2,warm} \approx 1\times10^{41}$ \ergs.  For the nuclear region to sustain this luminosity of warm H$_2$ would require an AGN with a bolometric luminosity of $\gtrsim 2\times10^{44}$ \ergs, assuming first that $L_{X{\rm, AGN}} \gtrsim 100 L_{\rm H_2, warm}$, \citealt{ogle+10}; and $L_{\rm bol, AGN} \gtrsim 16 L_X$, \citealt{ho2008}), a factor of three larger than $L_{\rm FIR}$ ($7\times10^{43}$ \ergs; \citealt{Gil+07}).  If this warm H$_2$ is indeed part of the outflow, our estimate of the total mass of the outflow should increase by 40\%.  Adding in the contribution of the ionized and hot gas would increase the mass of the outflow further, thus we consider our estimate of \mouttot\ \msun\ to be a conservative lower limit.

\begin{figure*}[ht!]
\centering
\subfigure{\includegraphics[height=4in,clip,trim=0cm 0cm 0cm 0cm]{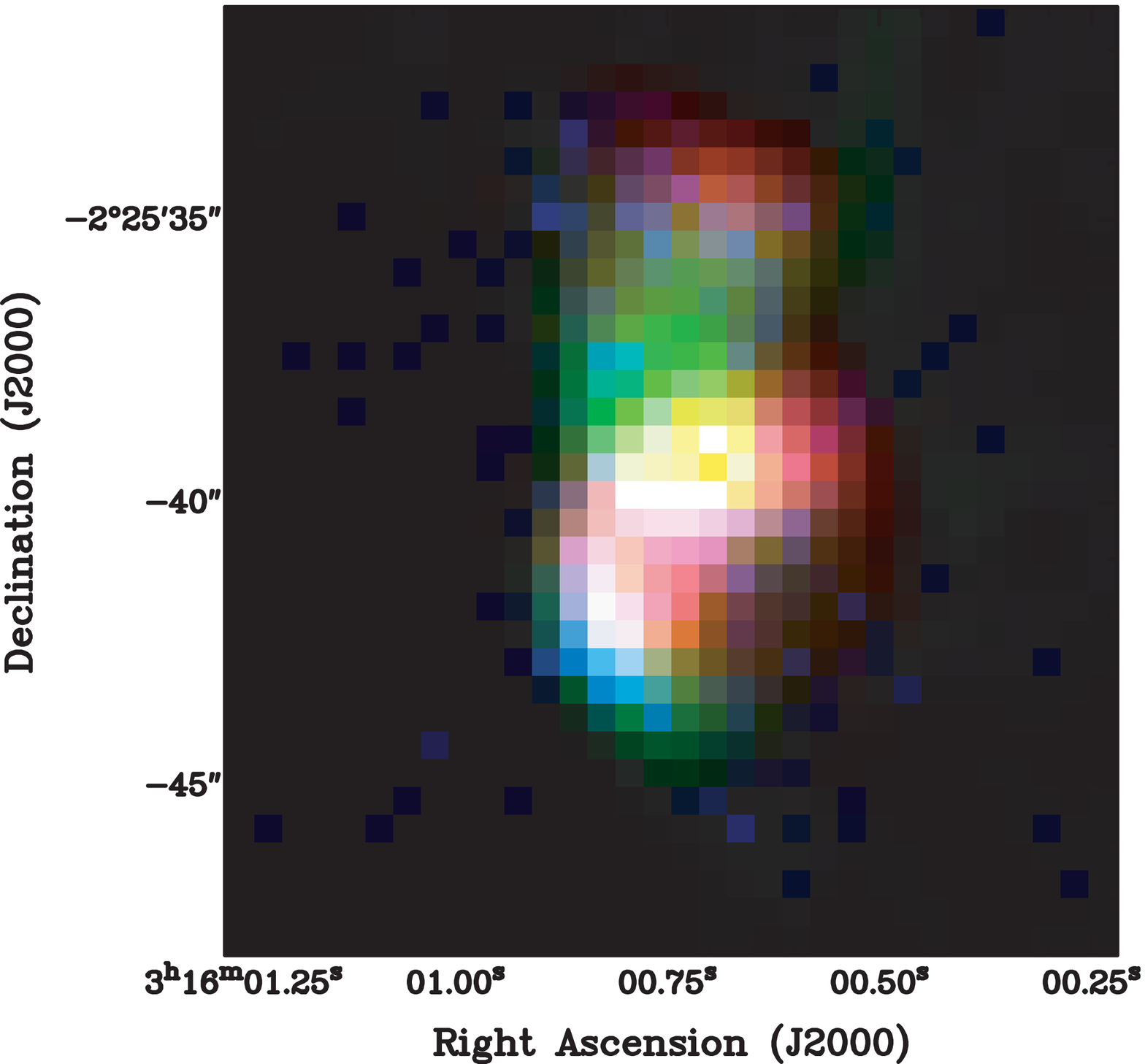} }
\subfigure{\includegraphics[height=4in,clip,trim=0.2cm 4.3cm 1.6cm 5.8cm]{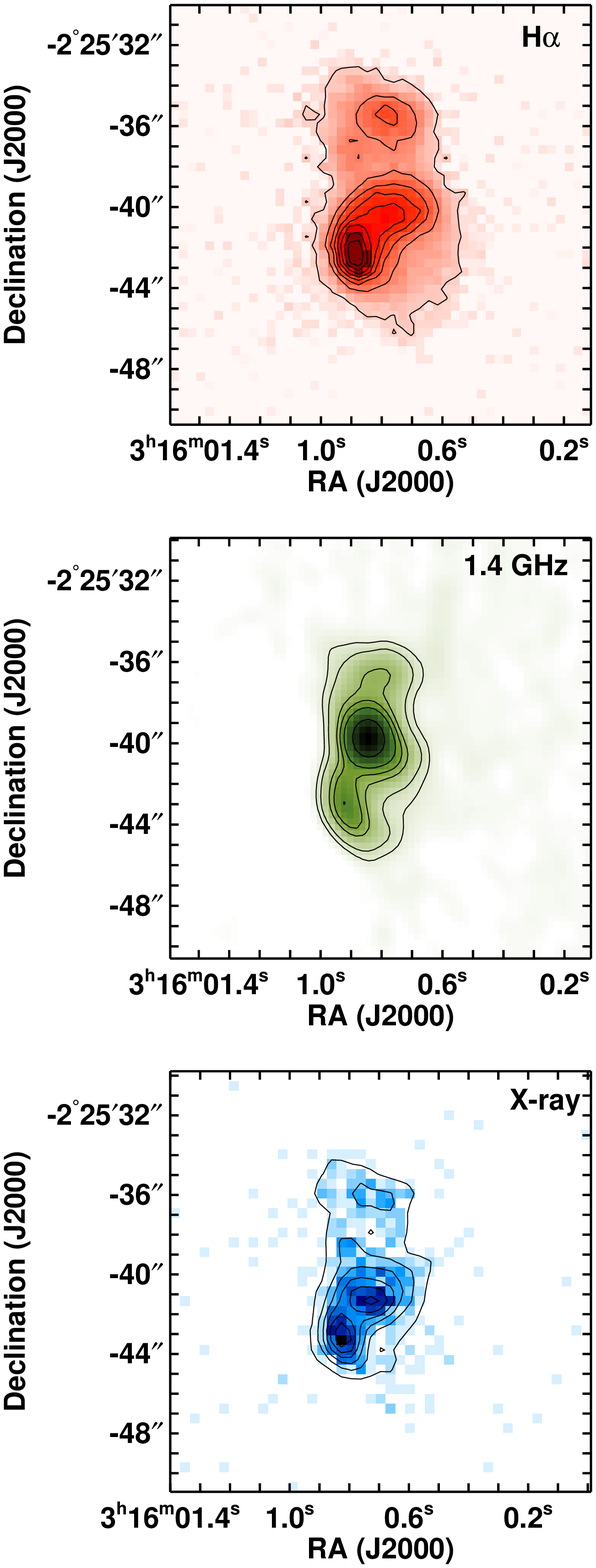}} 
\caption{{\bf (Left:)} A RGB image of the \ha\ (red), radio continuum (green), and {\it Chandra} X-ray (blue) emission.  A spatial correlation is clearly seen in the area of the blueshifted CO outflow, revealing that these features are co-spatial.  We conclude that the \ha, the spurs in the radio continuum image, and some of the X-ray emission originate from the outflowing material, although the 1.4 GHz continuum emission could also be coming from a radio jet. {\bf (Top right:)} \ha\ image from the SINGS survey.  {\bf (Middle right:)} VLA A-array 1.4 GHz continuum emission \citep{bk06}.  The unresolved peak in the emission is assumed to be from an AGN, and the lobe extending southeast to northwest is assumed to trace the interface between the outflowing material and the galaxy ISM.  {\bf (Bottom right:)} Un-smoothed {\it Chandra} X-ray image (Alatalo et al. 2011, in prep).  The majority of the X-ray photons from NGC~1266 can be fit with a thermal bremsstrahlung spectrum with an excess of hard X-rays.  All images are at the same spatial scale.}
\label{fig:shocks}
\end{figure*}

Combined with the molecular emission and \ion{H}{1} absorption data, X-ray, H$\alpha$, and radio continuum data further elucidate the nature of the outflow.  Figure \ref{fig:shocks} shows a comparison of the H$\alpha$ narrow-band image from the SINGS survey \citep{ken+03}, a 1.4 GHz radio continuum map \citep{bk06}, and the unsmoothed X-ray map from the {\it Chandra} X-ray Telescope Advanced CCD Imaging Spectrometer (ACIS) on the same spatial scale (Alatalo et al. 2011, in prep).  A spectral fit to the {\it Chandra} data co-added over the extent of the emission shows that the X-ray emission is dominated by thermal bremsstrahlung.  The 1.4 GHz continuum shows both a centrally peaked, unresolved core (possibly from a radio jet), as well as two more diffuse spurs running from the southeast to the northwest.  The lobed structure first apparent in the molecular gas (Fig. \ref{fig:co_mom1}) is seen in all these tracers.  In fact, the X-rays and the 1.4 GHz spurs, which do not suffer significantly from extinction, show an asymmetry between the foreground and background lobes, relative to the CO at $v_{\rm sys}$.  This likely means that the relative faintness of the \ha\ flux from the redshifted lobe as compared to the blueshifted lobe is not primarily due to extinction.  The spatial relationship between the molecular gas and the X-ray, 1.4 GHz and \ha\ emitting material supports this picture.  The \ha, radio spurs, and X-rays likely are emitted from the interface where the outflowing material collides with the interstellar medium (ISM) of NGC~1266.  This interpretation is supported by the thermal bremsstrahlung emission in the {\it Chandra} spectrum, although the radio spur emission could also possibly come from a radio jet.

The geometry of the CO outflow can be explained using two possible scenarios.  Firstly, it could be due to a molecular outflow launched deep in the nucleus of the galaxy, confined azimuthally by the massive molecular disk.  Secondly, it could directly originate from the molecular disk, and be entrained by hot winds launched by the AGN, surviving only because of the extreme density of the molecular disk. In either of these scenarios, at least some of the molecular material escapes from the galaxy still intact, and will energize the intergalactic medium (IGM). This is especially true if the outflow continues to accelerate from its launch point.  


\begin{figure}[h!] 
\centering
\includegraphics[width=3.0in,clip,trim=1.8cm 1.5cm 2.1cm 3.7cm]{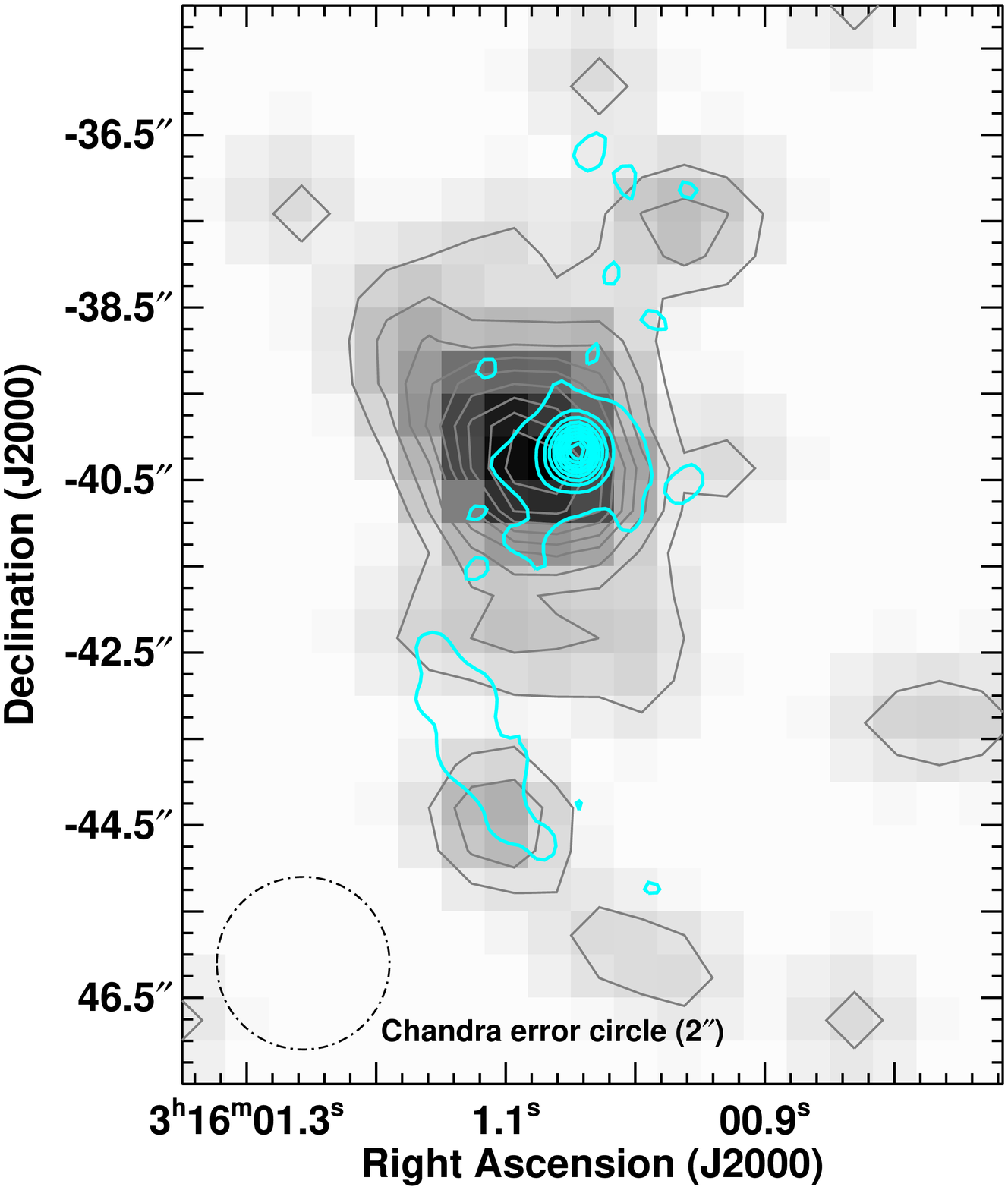}
\caption{{\it Chandra} smoothed hard X-ray image truncated to only include 4--8 keV photons (greyscale and black contours), overlaid with the 5 GHz radio continuum emission (cyan contours; \citealt{bk06}).  The radio core and hard X-rays from {\it Chandra} are co-spatial within the expected positional accuracy of the {\it Chandra} data in the hard X-ray band (2\arcsec), convincingly identifying NGC~1266 as harboring an AGN.}
\label{fig:chandra_5ghz}
\end{figure}

\subsection{Derived Properties}

A summary of the derived properties of NGC~1266 is presented in Table \ref{tab:results}. The mean molecular hydrogen surface density of the nucleus is \signuc\ \msun\ pc$^{-2}$, corresponding to a mean column density $ N({\rm H_2}) \approx $ \nhnuc\ cm$^{-2}$, two orders of magnitude higher than the surface density of individual GMCs in the Milky Way, and comparable to what is seen in some of the most luminous starburst galaxies (Kennicutt 1998).  

For an outflow with a mass of \mouttot\ \msun\ and an outflow velocity of half the full width-half maximum ($v_{\rm fwhm}$) of the broad wing component (177 \kms), the kinetic energy associated with the outflow is \kinenrg\ ergs, equivalent to the total kinetic energy expelled from \kesne\ supernova explosions.  The dynamical time, $\tau_{\rm dyn}$, for the outflow to reach its maximum radius of \rout\ pc at this velocity is only \tdyn\ Myr. If the edge of the CO emission in the lobes is produced from only the highest velocity gas, the relevant timescale is even shorter, $\tau_{\rm dyn} \sim$ \tdynex\ Myr.

From the total kinetic energy and $\tau_{\rm dyn}$, the mechanical luminosity of the outflow $L_{\rm outflow} = $ \lumout\ \ergs.  The mass outflow rate is $\dot{M} = M_{\rm outflow}/\tau_{\rm dyn}$ = \mflux\ \msunyr.  At this rate, if the source of the gas in the outflow is the nuclear molecular disk, the nucleus of NGC~1266 will be completely depleted in $\tau_{\rm dep} = $ \tdepnuc\ Myr, or \tdep\ Myr for all of the gas in the center (nucleus + envelope), both very short timescales.


\begin{table}[hb!]
\centering
\caption{NGC1266 Characteristic Properties}
\begin{tabular*}{7cm}{r l}
\hline \hline
$M_{\rm nucleus}$ & \mnuc\ \msun \\
$R_{\rm nucleus}$ & \rnuc\ pc\\
$M_{\rm CVC}$ & \mcvc\ \msun \\
$<\Sigma(\rm H_2)_{\rm nucleus}>$& \signuc\ \msun\ pc$^{-2}$\\
$<N(\rm H_2)_{\rm nucleus}>$  & \nhnuc\ cm$^{-2}$\\
$<n(\rm H_2)_{\rm nucleus}>$ & \nnuc\ cm$^{-3}$\\
\hline
$M_{\rm H~I+H_2, outflow}$ & \mouttot\ \msun \\
$M_{\rm H_2, outflow}$ & \mout\ \msun \\
$R_{\rm outflow}$ & 450 pc\\
$v_{\rm outflow}$ & 177 \kms\\
$\tau_{\rm dyn}$ & \tdyn\ Myr \\
$\dot{M}$ & \mflux\ \msunyr \\
${\rm KE_{outflow}}$ & \kinenrg\ ergs\\
$L_{\rm outflow}$ & \lumout\ \ergs\\
$\tau_{\rm dep}$ & \tdep\ Myr\\
\hline \hline
\end{tabular*}
\label{tab:results}
\end{table}

\section{Discussion}
\label{disc}

\subsection{Uniqueness of NGC~1266}
NGC~1266 is unusual in several respects.  There is a large reservoir of molecular gas concentrated within the central 100 pc.   The concentration of H$_2$ in the nucleus implies that the gas must have lost nearly all of its angular momentum.  Although such high central concentrations of H$_2$ are observed in interacting galaxies, NGC~1266 is an isolated galaxy with no sign of interaction or merger.



Is the outflow in NGC~1266 a common evolutionary stage through which most ETGs pass, or is NGC~1266 a pathological phenomenon? The short gas depletion time of $\approx$ \tdep\ Myr is reasonably consistent with finding only one such case so far in the 60 \atlas\ CO detections \citep{young+11}. The expectation would be to find \howmany, assuming that the typical molecular gas depletion time due to star formation in ETGs is the same as the 2 Gyr found in late-type galaxies \citep{leroy+08,bigiel+09}.  However, the sheer strength of the CO line in NGC~1266 certainly aided the outflow detection, and it is likely that, given the lower signal-to-noise ratios present in the rest of the sample, the low surface brightness features at high velocities would generally be lost in the noise.


Optically, NGC~1266 does not stand out within the \atlas\ sample.  It is a fast rotator with regular stellar kinematics \citep{kraj+11}, and is of normal metallicity.  Its absolute magnitude, $M_K = -22.9$, corresponds to a stellar mass just below the median of the sample.  On the other hand, its radio flux at 1.4 GHz places NGC~1266 in the top 10\% of radio emitters within the sample, although this emission could be explained by the FIR-radio SF correlation based on the amount of molecular gas detected in the source \citep{murgia+05}.  Among the $\sim100$ known ETGs with detected molecular gas (\citealt{wch95,ws03,swy07,cyb07}; Young et al. 2011),  $\approx 20$ have been previously mapped interferometrically (see \citealt{crocker+11}, and references therein).  Compared to those, the molecular gas in NGC~1266 stands out both in terms of its compactness and its large mass.  The star-forming ETGs seem to exhibit only a narrow range of mean molecular gas surface densities, 10--500 \msun\ pc$^{-2}$, similar to the range of Milky Way values.  The surface density of the nucleus of NGC~1266 ($\approx$ \signuc\ \msun\ pc$^{-2}$) is two orders of magnitude larger than this.

NGC~1266 should perhaps be compared to radio galaxies, 1/4 - 1/3 of which have been shown to contain significant reservoirs of molecular gas \citep{evans+05, ogle+10}. The \ion{H}{1} absorption profile seen in NGC~1266 is akin to that in the handful of known radio galaxies with outflowing \ion{H}{1}: IC~5063, 3C~305 \citep{morganti+98,oosterloo+00,morganti+05a,morganti+05b}, 4C~12.50 \citep{morganti+04} and 3C~293 \citep{morganti+03,emonts+05}, with mass outflow rates $\sim 50$ \msunyr\ at high ($\gtrsim 1000$ \kms) velocities.  The \ion{H}{1} in each of these galaxies is detected in absorption against a radio source, and covers hundreds of parsecs to kiloparsecs.  While the outflowing \ion{H}{1} mass in NGC~1266 is orders of magnitude lower than in these galaxies, it is possible that both types of systems share a common origin.  However, although the luminosity and structure of the \ion{H}{1} absorption do mirror those in radio galaxies, the samples described by \citet{evans+05} and \citet{ogle+10} and the outflowing systems all show signs of interactions, companions and tidal features, which are all markedly absent from NGC~1266.

\subsection{Star Formation in NGC~1266}

Given the high surface density of the molecular gas, NGC~1266 ought to be forming stars.  Based on the Kennicutt-Schmidt (K-S) relation \citep{Ken98}, $\Sigma_{\rm H_2} \approx$ \signuc\ \msun\ pc$^{-2}$, and a radius of \rnuc\ pc, the expected star formation rate (SFR) of the nucleus alone is \sfrgas\ \msunyr, adjusted for a Kroupa initial mass function (IMF).  In order to understand the star formation efficiency of the gas, we use multiple tracers to place constraints on the SFR, including the total far-infrared luminosity ($L_{\rm FIR}$), the 24\micron\ emission, and the 8\micron\ emission from polycyclic aromatic hydrocarbons (PAHs).  Those three tracers are systematically affected by the presence of an AGN, but in different ways, so we use all three to define a range of possible SFRs.  Both the 24\micron\ emission and $L_{\rm FIR}$ suffer from contamination by a buried AGN, and will thus systematically overestimate the SFR.  PAH molecules are destroyed by the harsh ionization fields associated with strong AGN, and will thus possibly underestimate the true SFR \citep{voit92}.

If we assume that all of the far-infrared (FIR) luminosity ($7\times10^{43}$ \ergs; \citealt{Gil+07,cyb07}) is from star formation (an upper limit since the nucleus of NGC~1266 contains an AGN), we calculate a global obscured SFR$_{\rm FIR}$ of 2.2 \msunyr\ using \citet{Ken98}, re-calibrated for a Kroupa IMF.  The unobscured SFR from the measured far ultraviolet (FUV) contributes negligibly to the overall SFR (SFR$_{\rm FUV} = 0.003$ \msunyr; \citealt{Gil+07,leroy+08}).  We also calculate the SFR using the 24\micron\ and PAH emission, measured by {\em Spitzer} as part of the SINGs survey \citep{ken+03}.  The SFR is related to the 24\micron\ luminosity ($L_{24}$) following \citet{calzetti+07}: SFR$_{24}$ / (\msunyr) $= 1.27\times10^{-38} (L_{24}/$\ergs$)^{0.885}$.  For NGC~1266, $L_{24} = 1.06\times10^{43}$ \ergs\ once adjusted to our adopted distance \citep{temi+09}, so the associated SFR$_{24} \approx 1.5$ \msunyr\ (the stellar continuum contribution to the 24\micron\ emission is $<< 1$\%), within 33\% of the SFR$_{\rm FIR}$ derived from the FIR emission.  To measure the SFR from the PAH emission, we use the mid-IR fluxes from Falc\'{o}n-Barroso et al. (in prep).  We subtract off the stellar contribution from the 8.0\micron\ flux by scaling the 3.6\micron\ image; the scale factor of the 3.6\micron\ image can range from 0.22 to 0.29 \citep{calzetti+07}, with elliptical galaxies having typical values of 0.26 \citep{wu+05, shapiro+10}.  From the corrected PAH emission, we measure SFR$_{\rm PAH} \approx 0.57$ \msunyr (Falc\'{o}n-Barroso et al. 2011, in prep) a factor of 4 lower than the SFR$_{\rm FIR}$ derived using the FIR emission.  

Comparing the SFR predicted from the molecular gas via the K-S relation to the SFRs derived from the 8\micron, 24\micron\ and FIR emission, we find that the predicted SFR is a factor of 1.4 to 5 times larger than the measured SFR, depending on which SFR tracer is used.  However, NGC~1266 is well within the scatter of the K-S relation \citep{leroy+08}, as well as the range of SFRs observed in the {\tt SAURON} and \atlas\ sample galaxies (\citealt{shapiro+10}; Falc\'{o}n-Barroso et al. in prep).

The star formation-driven wind model of \citet{Murray+05} requires that the star formation rate be equal to or greater than the mass outflow rate for radiation-driven outflows, which we refer to as the Murray criterion.  Our calculations for NGC~1266 show that even if we consider the higher SFR estimate (${\rm SFR} \approx 2$ \msunyr), our lower limit on the mass outflow rate ($\dot{M} \approx$ \mflux\ \msunyr) still exceeds it by a factor of at least a few.  This will only increase with more accurate accounting of the the total mass of the outflow.  Therefore, the NGC~1266 molecular outflow does not fulfill the Murray criterion, and momentum coupling (i.e. radiation-driving) and star formation are unlikely to be the drivers of the outflow.  

\subsection{AGN Feedback and Comparison to Other Molecular Outflows}
\label{othersys}

NGC~1266 is classified as a low-ionization nuclear emission-line region (LINER) \citep{moustakas+06}.  The 1.4 and 5 GHz radio continuum images reveal a prominent and unresolved point source in the nucleus (e.g. Figs \ref{fig:shocks} and \ref{fig:chandra_5ghz}).  {\it Chandra} observations reveal hard X-ray emission centered on the radio point source, consistent with an obscured AGN (see Figure \ref{fig:chandra_5ghz}).  Given the overwhelming evidence for the presence of an AGN in NGC~1266, and that star formation is much less intense than in the few other galaxies exhibiting molecular outflows (see \S\ref{othersys}), an AGN-driven outflow seems a reasonable possibility.

To investigate whether the AGN is any more capable, we use the relationship between the 1.4 GHz luminosity and total jet power from \citet{birzan+08}: $\log (P_{\rm jet}/10^{42}$ \ergs$) = 0.35 \log (P_{\rm 1.4GHz}/10^{24}~{\rm W~Hz^{-1}}) + 1.85$.  The total radio flux at 1.4 GHz is $9.3\times 10^{20}$ W Hz$^{-1}$, resulting in a total jet power of $6.1\times10^{42}$ \ergs.  Comparing this to the $L_{\rm mech}$ of the outflow of \lumout\ \ergs, a coupling of $\sim2$\% is required to drive the outflow with the AGN, a rather modest value.  Therefore, the AGN does indeed generate sufficient energy to power the outflow through the mechanical work of the radio jet.

There is a small sample of galaxies that have been shown to exhibit molecular outflows: M82, Arp~220, Mrk~231 and M51.  Arp~220 is a starbursting major merger and has a molecular gas surface density of 6 $\times\ 10^4$ \msun~pc$^{-2}$ \citep{mauer+96}, similar to that of the molecular nucleus of NGC~1266.  \citet{sakamoto+09} detected an outflow that they estimate has a mass of 5 $\times\ 10^7$ \msun\ and a mass outflow rate of 100 \msunyr, with an outflow velocity of $\sim100$ \kms. These numbers, though uncertain, are also comparable to what is observed in NGC~1266.  However, the star formation rate in Arp 220 is $\gtrsim 200$ \msunyr~\citep{soifer+87}, almost two orders of magnitude greater than in NGC~1266, providing a ready source of energy for the outflow and meaning that the Murray criterion is met.
   
M82 has $\approx3$ $\times\ 10^8$ \msun\ of molecular gas entrained in an outflow with velocities up to $\approx230$ \kms\ \citep{walter+02L}. It is unclear whether this is sufficient for any gas to escape the galaxy.  The mass outflow rate is $\approx 30$ \msunyr, comparable to that estimated for the wind in NGC~1266, and roughly equivalent to the global SFR of M82 \citep{Ken98}, thus satisfying the Murray criterion.  The outflows can thus be driven out via radiation pressure.

\citet{matsu+07} have shown recently that M51 has entrained molecular gas in an outflow close to the AGN, with a total molecular mass of 6 $\times\ 10^5$ \msun, two orders of magnitude less than the mass estimate for the NGC~1266 outflow.  The derived mass outflow rate of M51 is $\approx 4$ \msunyr.  While this outflow rate is equivalent to the total SFR, the star formation in M51 is known to be distributed throughout the disk \citep{calzetti+05}, and thus is likely unable to effectively drive much of the nuclear molecular gas out.  The AGN luminosity of $2\times10^{42}$ \ergs\ \citep{tera+wilson01} is also too low to sustain the molecular outflow via photon-driving.  Therefore, M51 appears to be a scaled down version of the outflow we observe in NGC~1266, requiring a similar mechanism, such as mechanical work done by a radio jet, to drive out the gas.

Mrk 231, a nearby advanced major merger, shows a molecular outflow similar to that of NGC~1266 \citep{feruglio+10}.  The Mrk 231 \hbox{CO(1--0)} spectrum exhibits a similar profile to that seen in NGC~1266, with broad wings requiring a nested Gaussian fit.  This similarity seems to confirm that neither NGC~1266 nor Mrk 231 is a unique case, and AGN feedback may be an efficient means of removing the molecular gas close to the AGN.  \citet{feruglio+10} claim that Mrk~231 has a mass outflow rate of $\approx 700$ \msunyr\ (assuming a density profile of $r^{-2}$), exceeding its star formation rate of 200 \msunyr and thus requiring AGN feedback to be powered.  If the CO in the outflow is optically thick, as assumed by \citet{feruglio+10}, then the mass outflow rate in Mrk~231 is indeed too high for a star formation-driven wind.  On the other hand, if the correct $L$(CO)-to-M(H$_2$) mass conversion for both the NGC~1266 and Mrk~231 outflows is the optically thin conversion (as assumed for NGC~1266), then $M$(H$_2$) for the Mrk~231 outflow is a factor of $\approx 5$ lower, leading to a mass outflow rate of $\sim 100$ \msun\ yr$^{-1}$.  This is sustainable using the Murray criterion, and is similar to Arp 220.  Therefore the case for the molecular outflow in Mrk~231 requiring AGN feedback is uncertain.  To obtain a robust estimate of the outflowing mass of Mrk~231, the opacity of the outflowing component must be determined.

New stacked CO(1--0) observations of local ultra-luminous infrared galaxies (ULIRGs) by \citet{chung+10} also show clear evidence that high-velocity broad emission at up to 1000 \kms\ from the systemic velocity is a common feature in these systems.  In particular, the authors find that only the starburst--dominated ULIRGs exhibit broad wings, while AGN--dominated ULIRGS (identified by their optical spectra) do not.  These results are thus clear evidence that in most observed molecular outflows, photon-coupling with the star formation is the principle driver.

While the aforementioned galaxies all exhibit molecular outflows, 
M51 is the only other bonafide case of a non-photon-driven molecular outflow.  However, it has almost two orders of magnitude less molecular gas being driven out than NGC~1266.  Mrk~231 might well be an equivalent system, but the mechanisms capable of driving the gas out depends on the CO luminosity--to--H$_2$ mass conversion in the outflow.  The most acute contrast between NGC~1266 and the other molecular outflow sources is its lack of interaction.  All other galaxies are undergoing interactions; major mergers in the case of Arp~220, Mrk~231 and some of the \citep{chung+10} ULIRGs.  NGC~1266 shows no evidence of having undergone an interaction at any time in the recent past, with unperturbed stellar kinematics, no companion, and a lack of \ion{H}{1} emission internal or external to the galaxy.  This means that while we can point to an interaction as the triggering mechanism in the other known molecular outflows, the striking absence of such a perturbing event makes NGC~1266 even more remarkable.

\section{Summary and Conclusions}
\label{conclu}

We report the detection of a massive centrally-concentrated molecular component ($M_{\rm CVC} = $ \mcvc\ \msun) and a powerful molecular outflow ($M_{\rm H_2,outflow} =$ \mout\ \msun) in the field S0 galaxy NGC~1266.  The maximum velocity of the wind exceeds the escape velocity so that at least some gas will escape the galaxy to energize the IGM, and the total neutral gas mass outflow rate $\dot{M} \approx$ \mflux\ \msunyr.  The star formation rate in NGC~1266 appears to be incapable of driving this outflow; the AGN appears to be the main driving mechanism, as for \ion{H}{1} outflows observed in radio galaxies.  It is however unclear how the gas lost its angular momentum to fall so completely into the nucleus.

The central molecular gas contains a rapidly rotating nuclear disk of $\approx$ \rnuc\ pc radius enshrouded in a diffuse molecular envelope, and the molecular outflow emerges normal to the disk plane.  Although the molecular nucleus is compact and very near the AGN, multiple estimates of the star formation rate point to a low star formation efficiency, but the uncertainties in these SFRs place NGC~1266 within the scatter of the K-S relation (assuming that L$_{\rm FIR}$ is not dominated by the AGN).

If the gas in the nucleus is the source of the molecular outflow, the gas depletion timescale is \tdep\ Myr, short enough for all ETGs to go through such a phase while remaining consistent with having found only one such case among the 260 ETGs in the \atlas\ sample.  NGC~1266 is the brightest detection, however, so there might be more outflows in the sample, undetected so far because of lower S/N ratios.

While molecular outflows have been identified in a handful of galaxies, NGC~1266 is the only one that shows no evidence of having undergone an interaction, leaving the mechanism to transport the molecular gas into the center unknown.

\acknowledgments{{\it Acknowledgments:}  KA would like to thank P. Chang, C. Heiles, N. Murray and R. Plambeck for useful ideas and conversations.  KA would also like to thank A. Chung, M. Yun and C. Feruglio and the referee for their helpful discussions.  The research of KA is supported by the NSF grant AST-0838258.  RLD is supported by the rolling grants `Astrophysics at Oxford' PP/E001114/1 and ST/H002456/1 from the UK Research Councils. RLD acknowledges travel and computer grants from Christ Church, Oxford and support from the Royal Society in the form of a Wolfson Merit Award 502011.K502/jd.  TN acknowledges support from the DFG Cluster of Excellence: "Origin and Structure of the Universe."  Support for this work was provided by the National Aeronautics and Space Administration through Chandra Award Number 11700538 issued by the Chandra X-ray Observatory Center, which is operated by the Smithsonian Astrophysical Observatory for and on behalf of the National Aeronautics Space Administration under contract NAS8-03060.  Support for CARMA construction was derived from the states of California, Illinois, and Maryland, the Gordon and Betty Moore Foundation, the Kenneth T. and Eileen L. Norris Foundation, the Associates of the California Institute of Technology, and the National Science Foundation. Ongoing CARMA development and operations are supported by the National Science Foundation under a cooperative agreement, and by the CARMA partner universities.  The Submillimeter Array is a joint project between the Smithsonian Astrophysical Observatory and the Academia Sinica Institute of Astronomy and Astrophysics and is funded by the Smithsonian Institution and the Academia Sinica.  This research has made use of data obtained from the Chandra Data Archive and the Chandra Source Catalog, and software provided by the Chandra X-ray Center (CXC) in the application packages CIAO, ChIPS, and Sherpa.  The National Radio Astronomy Observatory is a facility of the National Science Foundation operated under cooperative agreement by Associated Universities, Inc.  This paper is partly based on observations carried out with the IRAM 30m telescope. IRAM is supported by INSU/CNRS (France), MPG (Germany) and ING (Spain).  We acknowledge use of the HYPERLEDA database (http://leda.univ-lyon1.fr) and the NASA/IPAC Extragalactic Database (NED) which is operated by the Jet Propulsion Laboratory, California Institute of Technology, under contract with the National Aeronautics and Space Administration.  MC acknowledges support from a STFC Advanced Fellowship (PP/D005574/1) and a Royal Society University Research Fellowship.  RMcD is supported by the Gemini Observatory, which is operated by the Association of Universities for Research in Astronomy, Inc., on behalf of the international Gemini partnership of Argentina, Australia, Brazil, Canada, Chile, the United Kingdom, and the United States of America.  
\\
\bibliographystyle{apj}
\bibliography{ms}

 \end{document}